\documentclass[opre,nonblindrev]{informs3} 

\DoubleSpacedXI 

\usepackage{endnotes}
\let\footnote=\endnote
\let\enotesize=\normalsize
\def\notesname{Endnotes}%
\def\makeenmark{$^{\theenmark}$}
\def\enoteformat{\rightskip0pt\leftskip0pt\parindent=1.75em
  \leavevmode\llap{\theenmark.\enskip}}

\usepackage{natbib}
 \bibpunct[, ]{(}{)}{,}{a}{}{,}%
 \def\bibfont{\small}%
\TheoremsNumberedThrough     
\ECRepeatTheorems

\EquationsNumberedThrough    

\usepackage{adjustbox}
\usepackage{appendix}
\usepackage{multirow}
\usepackage{booktabs}
\usepackage{graphicx}
\usepackage{enumitem}
\usepackage{subcaption}
\usepackage{comment}
\usepackage{amsmath,amsfonts,amssymb,bbm,bm}
\usepackage{pgfplots}
\pgfplotsset{compat=1.16}
\usetikzlibrary{positioning}
\def\RP{{\mathbb{R}}_{\geq 0}}

\def\RPP{{\mathbb{R}}_{>0}}

\def\CG{{\sf CG}}
\def\LBG{{\sf LBG}}
\def\NCG{{\sf NCG}}
\def\PNCG{{\sf PNCG}}
\def\PLG{{\sf PLG}}
\def\2PLG{{\sf 2PLG}}
\def\PNC2{{\sf 2PNCG}}
\def\N{[n]}

\def\poa{{\sf{PoA}}}

\newcommand{\sg}{{\bm\sigma}}

\allowdisplaybreaks

\newtheorem{fact}[theorem]{Fact}

\usepackage[colorinlistoftodos,textsize=tiny,textwidth=35pt]{todonotes}
\usepackage{xcolor}
\newcommand\FB[1]{{\color{orange} #1}}

\begin{document}


\RUNAUTHOR{Benita et al.}

\RUNTITLE{Data-Driven Models of Selfish Routing}

\TITLE{Data-Driven Models of Selfish Routing:  Why Price of Anarchy Does Depend on Network Topology\endnote{A preliminary version of this work appeared in the proceedings of the 16th Conference on Web and Internet Economics (WINE 2020) \citep{BenitaBMPV20}.}}

\ARTICLEAUTHORS{%
\AUTHOR{Francisco Benita}
\AFF{Singapore University of Technology and Design, \EMAIL{francisco\_benita@sutd.edu.sg}} 
\AUTHOR{Vittorio Bil\`o}
\AFF{University of Salento, \EMAIL{vittorio.bilo@unisalento.it}}
\AUTHOR{Barnab\'e Monnot}
\AFF{Ethereum Foundation, \EMAIL{barnabemonnot@gmail.com}}
\AUTHOR{Georgios Piliouras}
\AFF{Singapore University of Technology and Design, \EMAIL{georgios@sutd.edu.sg}}
\AUTHOR{Cosimo Vinci}
\AFF{University of Salerno, \EMAIL{cvinci@unisa.it}}
\AFF{Gran Sasso Science Institute\endnote{The current affiliation of Cosimo Vinci is the University of Salerno. A substantial part of this work was done when he was affiliated with the Gran Sasso Science Institute (that is his previous affiliation).}}
} 

\ABSTRACT{
We investigate traffic routing both from the perspective of theory as well as real world data. First, we introduce a new type of games: $\theta$-free flow games. Here, commuters only consider, in their strategy sets, paths whose free-flow costs (informally their lengths) are within a small multiplicative $(1+\theta)$ constant of the optimal free-flow cost path connecting their source and destination, where $\theta\geq0$. We provide an exhaustive analysis of tight bounds on PoA($\theta$) for arbitrary classes of cost functions, both in the case of general congestion/routing games as well as in the special case of path-disjoint networks. Second, by using a large mobility
dataset in Singapore, we inspect minute-by-minute decision-making of thousands of commuters, and find that $\theta=1$ is a good estimate of agents' route (pre)selection mechanism. In contrast, in Pigou networks, the ratio of the free-flow costs of the routes, and thus $\theta$, is \textit{infinite}; so, although such worst case networks are mathematically simple, they correspond to artificial routing scenarios with little resemblance to real world conditions, opening the possibility of proving much stronger Price of Anarchy guarantees  by explicitly studying their dependency on $\theta$. For example, in the case of the standard Bureau of Public Roads (BPR) cost model, where$c_e(x)= a_e x^4+b_e$, and for quartic cost functions in general, the standard PoA bound for $\theta=\infty$ is $2.1505$, and this is tight both for  general networks  as well as path-disjoint and even parallel-edge  networks. In comparison, for $\theta=1$, the PoA in the case of general networks is only $1.6994$, whereas for path-disjoint/parallel-edge networks is even smaller ($1.3652$), showing that both the route geometries as captured by the  parameter $\theta$ as well as the network topology have significant effects on PoA. 

}


\maketitle

%


\section{Introduction}
\label{sec:intro}

Modern cities are wonders of emergent, largely self-organizing, behavior. Major capitals buzz with the collective hum of millions of people whose lives are intertwined and coupled in myriad and diverse ways. One of the most palpable such phenomena  of collective behavior is the emergence and diffusion of traffic throughout the city. A bird's eye view of any major city would reveal a complex and heterogeneous landscape of thousands upon thousands of cars, buses, trucks, motorcycles, running though the veins of a maze of remarkable complexity and scale consisting of a vast number of streets and highways.  The full magnitude of the multi-scale complexity of these real-life networks lies outside the perceptive capabilities  of any single individual.  Nevertheless, as a phenomenon that we get to experience daily, such as the weather, we would like to understand at least some macroscopic, high level characteristics of traffic routing. Quite possibly, one of the most interesting such questions is how efficient is a traffic network?

This question has received a lot of attention within algorithmic game theory.  Using the model of congestion games, seminal papers in the area established tight bounds on their Price of Anarchy (PoA), i.e., the worst case inefficiency of traffic routing \citep{KoutsoupiasP99WorstCE,roughgarden2002bad}. For example, the Price of Anarchy  of linear non-atomic congestion games is $4/3$, whereas if we apply the standard Bureau of Public Roads (BPR) cost functions that are polynomials of degree four, then the Price of Anarchy is roughly $2.151$. On the positive side, these bounds apply to all networks (within the prescribed class of delay/cost functions) regardless of their size or their total demand, or number of agents and are tight even for the simplest possible network instances, i.e., Pigou networks with just two parallel links. 



The common interpretation of these bounds is that they are strong and a PoA  anywhere in that range (e.g. PoA$=2$)  immediately translates to practical guarantees about real traffic. Some recent purely experimental work, however, has produced new insights that allow us to reexamine these results from a different perspective. For example, \cite{monnot2017routing} showed that the efficiency of real-life traffic networks, as estimated from traffic measurements alone, is really close to optimal even when compared to very optimistic estimates of optimal performance. A Price of Anarchy of $2$ implies that the average commuter can 
 increase their mean speed by $100\%$. Measurements suggest that this level of inefficiencies/improvements is rather unlikely.
Since Price of Anarchy is a macroscopic characteristic of a system with countless moving parts, a more useful analogy is that of weather or climate (e.g., average temperature). The differences between $10\%$ and $20\%$  increase to system inefficiency are  significant and a 100\% increase, i.e., PoA of $2$ would have catastrophic consequences.

\textit{A natural question emerges:} Can we create classes of models, i.e., congestion games, which come closer to representing real world traffic? In this paper we do, by leveraging an intuitive but largely unexplored  characteristic of real world traffic routing. Commuters only consider in their strategy sets paths/routes whose free-flow costs (informally their lengths, or their costs in absence of congestion) are approximately equal to each other (within a multiplicative factor of $1+\theta$). We call such games $\theta$-free flow games. 
We generalize the  special case  of linear congestion $\theta$-free flow games \citep{BV20} to the case of arbitrary classes of cost functions and simultaneously study both general and path-disjoint networks. 
 $\theta=0$ means that all paths considered by each user have exactly equal free-flow cost/length, whereas $\theta=1$ allows for paths whose lengths are within a factor of $2$. Pigou networks may feel intuitively very simple and thus natural due to their small size, but they fail to satisfy this property in the most extreme sense. The ratio of the free-flow costs of the two edges is infinite ($\theta=\infty$), since the cost of one edge is unitary (independently on its congestion), while the other edge has a null cost in absence of congestion. It is like considering two possible paths from home to work, one which is the shortest distance route and one that circumnavigates the globe along the way. Such unnatural paths may indeed be available to us, but we unconsciously and automatically prune them out from the set of alternatives that we consider. Amazingly, enforcing such a natural property on the set of models (routing games) we consider immediately removes from consideration Pigou networks, the worst case examples from a PoA perspective, and thus opens up the possibility of proving stronger Price of Anarchy guarantees. What are the implications of such characteristics to PoA? What other type of attributes can we take advantage of when creating new models? Finally, how well do they match real traffic conditions?

\subsection{Our Contribution}
\label{sec:contributions}

\begin{figure}[h!]
\centering
\includegraphics[scale=0.60]{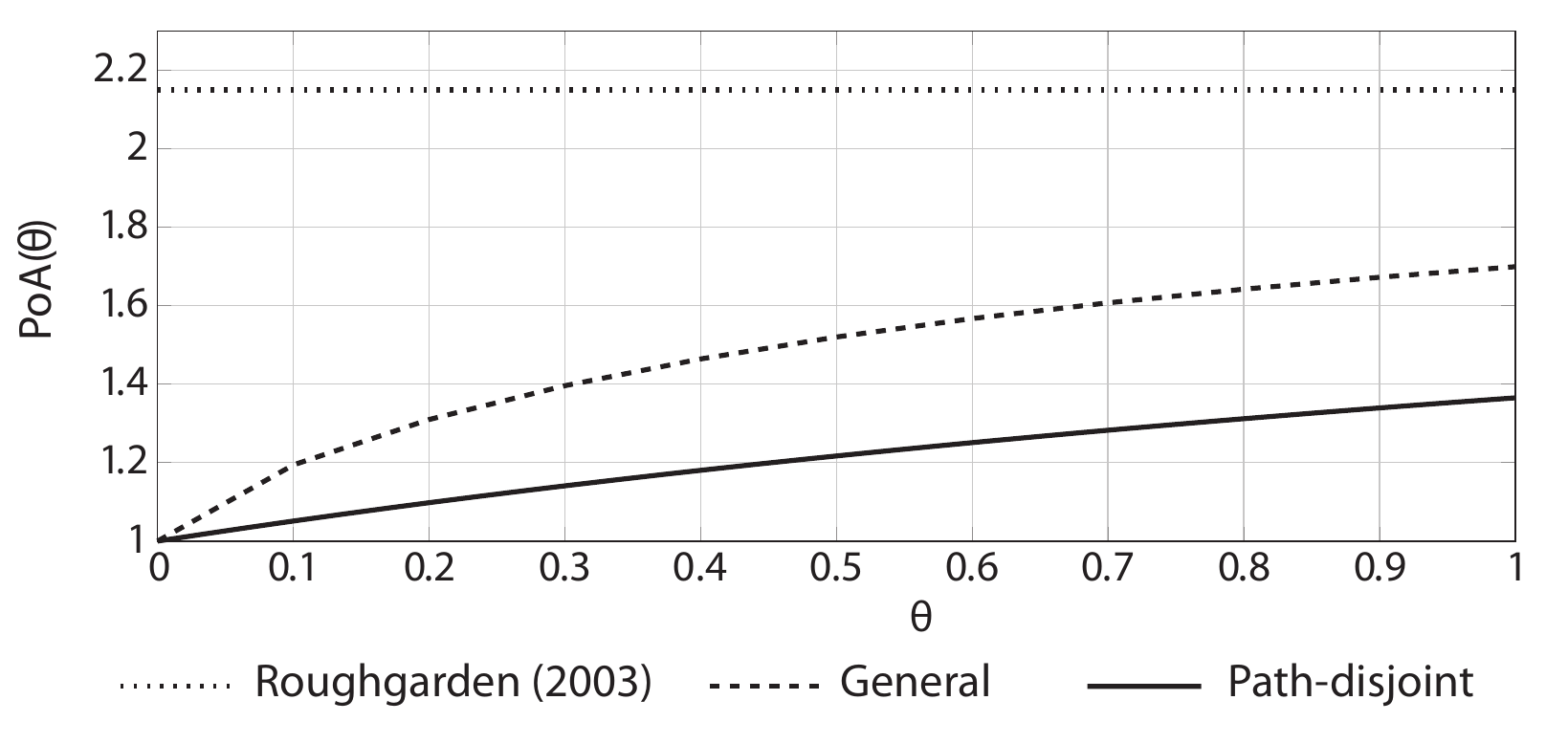}\caption{Comparison between PoA($\theta$) in the case of quartic costs for general/path-disjoint networks resp. and the standard bound PoA($\infty$)=$2.1505$ from \cite{R03}.}\label{Fig: main-a}
\end{figure}

In Section \ref{sec:model} we introduce a new class of congestion games, that we call free-flow games, parametrized by $\theta$. Building on the primal-dual method introduced in \cite{B18}, we provide two parametric {\em tight bounds} on the Price of Anarchy of free-flow games under general latency functions satisfying mild assumptions, thus largely extending the results given in \cite{BV20} which are restricted to affine latencies only. The first of these bounds applies to the general case of unrestricted network topologies (Theorem \ref{thm_gen_upp}) which includes single-source network congestion games and load balancing games, while the second one holds for path-disjoint networks (Theorem \ref{thm_gen_upp_par}) which includes the fundamental parallel-link topology. These bounds are equal if $\theta\in\{0,\infty\}$, but might be different if $\theta\in (0,\infty)$ (see Subsection \ref{gen_par_sub}). In fact, differently from what happens in the classical setting without the free-flow assumption, where the worst-case situation already arises in a two parallel-link network (the Pigou network), for free-flow games the absence of intersections among paths allows for more efficient equilibria. More precisely, as $\theta$ goes to infinity, both bounds converge to the same limit, but the convergence of the one for parallel-link networks can be significantly slower (see, for instance, Figure \ref{Fig: main-a}). We also stress that, with respect to the case of affine latency functions, our findings improve on the results given in \cite{BV20}, as we close the gap between upper and lower bound on the Price of Anarchy for parallel-link networks that was left as an open problem (see Subsection \ref{fre_pol_sub}).

In Section \ref{sec:data}, we  experimentally compute estimates of $\theta$ from real world traffic data. We employ an experimental dataset that contains detailed information (sampled every 13 seconds) on the routing behavior of tens of thousands of commuters in Singapore. Based on this fine-grained information and in combination with a graph  representation of the road network of Singapore that we have created we can estimate numerous characteristics of the actual routing behavior at an unprecedented level of accuracy. Using these tools that we believe are of independent interest as well, we find that the $\theta$ values for the vast majority of commuters (close to $80\%$) are below $1$.

One of the most important messages coming from our investigation is that the separation outlined by Theorems \ref{thm_gen_upp} and \ref{thm_gen_upp_par} sheds new light on the question of whether the Price of Anarchy is affected by the network topology. In fact, a famous, and perhaps counter-intuitive, result by Roughgarden \citep{R03} states that the PoA is independent of the network topology as,  in almost all notable cases, worst-case instances are already attained by simple networks, such as parallel-link graphs. Under the free-flow assumption, however, this situation ceases to hold, and the network topology begins to play a critical, if not dominant, role in the efficiency of equilibria. This evidence has major practical implications, as it signifies the fundamental importance of careful road network design and planning for selfish routing. As shown in Figure  \ref{Fig: main-a} and in more details in Table \ref{table1},
 in the case of the standard Bureau of Public Roads (BPR) cost model, $c_e(x)= a_e x^4+b_e$ and more generally quartic cost functions, applying the constraint
  $\theta=1$ nearly \textit{halves the percentage of inefficiency}, and applying the additional constraint of a path-disjoint network \textit{halves it once again}.


At the technical level, our general formulas depend on whether the free-flow traversing time of some edges is larger than zero, i.e., whether the limit of the edge cost/latency as its load goes to zero is strictly positive. Latency functions for which this does not hold have been termed {\em homogeneous} by  \cite{R03} and they represent one of the few exceptions for which he could not prove that the PoA is independent of the network topology. As a by-product of our results, we also obtain that, for games with homogeneous latency functions (which are a particular case of $0$-free-flow games), the Price of Anarchy is lower than the one attained by non-homogeneous latencies, and it is tight even for parallel-link topologies (see Corollary \ref{cor1} and Subsection \ref{sub_int_1}), thus answering the open question posed by \cite{R03}. 

To summarize, we obtain that the worst-case PoA is attained by parallel-link games if and only if one of the following cases occurs: (i) $\theta=0$ (which includes the case of homogeneous latency functions as a special case) and (ii) $\theta=\infty$.

For the sake of a more concrete exposition of our results and for empirical purposes, we provide explicitly an instantiation of the PoA bounds in the case of polynomial latency functions (Theorems \ref{thm_gen_pol} and \ref{thm_gen_pol_par}). The resulting bounds depend on both the maximum and minimum degree of the polynomials and, in the case of non-homogeneous polynomials only, they also depend on $\theta$. A quantitative representation of our results is partially summarized in Table \ref{table1}.

\begin{table}[!h]
\caption{The Price of Anarchy of free-flow games with non-homogeneous (i.e., with constant terms allowed) polynomial latency functions of maximum degree $p\leq 4$ and minimum degree $q$. Unlabelled bounds are proven in this paper. Bounds for homogeneous (i.e., without constant terms) polynomials can be obtained from the case $\theta=0$ (the same upper bounds have been given in \cite{DG06}, but tight lower bounds were only conjectured to exist). As it can be appreciated, the PoA depends on the network topology whenever $0<\theta<\infty$. Notes: a: \cite{BV20}; b: \cite{roughgarden2002bad}; c: \cite{R03}.}\label{table1}
\begin{adjustbox}{width=1\textwidth}
\begin{tabular}{|c||c|c||c|c||c|c||c|c|}
\hline
\multirow{2}{*}
{$(p,q)$} &
\multicolumn{2}{|c|} {$\theta = 0$} & \multicolumn{2}{|c|}{$\theta=1/2$} & \multicolumn{2}{|c|}{$\theta=1$} & \multicolumn{2}{|c|}{$\theta=\infty$}\\
\cline{2 - 9} & General & Path-disjoint & General & Path-disjoint & General & Path-disjoint & General & Path-disjoint\\
\hline\hline
$(1,1)$ & $1^a$  & $1$ & $1.1547^a$  & $1.0909$ & $1.2071^a$  & $1.1429$ & $1.3333^b$  & $1.3333^b$ \\
$(2,1)$ & $1.0355$ & $1.0355$ & $1.2873$ & $1.1472$ & $1.3852$ & $1.2383$ & $1.6258^c$  & $1.6258^c$ \\
$(2,2)$ & $1$ & $1$ & $1.2873$ & $1.1472$ & $1.3852$ & $1.2383$ & $1.6258^c$  & $1.6258^c$ \\
$(3,1)$ & $1.0982$ & $1.0982$ & $1.4078$ & $1.1869$ & $1.5475$ & $1.3093$ & $1.8956^c$  & $1.8956^c$ \\
$(3,2)$ & $1.0147$ & $1.0147$ & $1.4078$ & $1.1869$ & $1.5475$ & $1.3093$ & $1.8956^c$  & $1.8956^c$ \\
$(3,3)$ & $1$ & $1$ & $1.4078$ & $1.1869$ & $1.5475$ & $1.3093$ & $1.8956^c$  & $1.8956^c$ \\
$(4,1)$ & $1.1676$ & $1.1676$ & $1.5202$ & $1.2170$ & $1.6994$ & $1.3652$ & $2.1505^c$  & $2.1505^c$ \\
$(4,2)$ & $1.0450$ & $1.0450$ & $1.5202$ & $1.2170$ & $1.6994$ & $1.3652$ & $2.1505^c$  & $2.1505^c$ \\
$(4,3)$ & $1.0080$ & $1.0080$ & $1.5202$ & $1.2170$ & $1.6994$ & $1.3652$ & $2.1505^c$  & $2.1505^c$ \\
$(4,4)$ & $1$ & $1$ & $1.5202$ & $1.2170$ & $1.6994$ & $1.3652$ & $2.1505^c$  & $2.1505^c$ \\
\hline
\end{tabular}
\end{adjustbox}
\end{table}

\subsection{Related Work}

\textit{Price of Anarchy in routing games:}
Introduced by  \cite{KoutsoupiasP99WorstCE}, the ratio between the social cost of the worst equilibrium of a game and its optimum was given the name Price of Anarchy (PoA) in \citep{papadimitriou2001algorithms}.
For networks of linear latency and general topology, PoA was bounded tightly by 4/3 \citep{roughgarden2002bad} and 5/2 in the atomic case \citep{christodoulou2005price}.
\cite{roughgarden2015intrinsic} addressed more general latency functions in atomic routing games and again gave tight bounds on PoA. However, for a large class of natural latency functions, PoA tends to 1 as the demand on the network approaches infinitesimally small or infinitely high levels \citep{colini2017asymptotic,Colini2020}. 
This
casts doubts on the predictive power of PoA on the state of a real system, as noted in \cite{monnot2017routing} and \citep{Wu2021}. 

Other inefficiency metrics widely investigated in selfish routing are the Price of Stability \citep{AnshelevichDKTWR08,ChristodoulouKS11}, that compares the best equilibrium of a game with its optimum, and the Price of Risk Aversion \citep{LianeasNM19,KleerS19,FotakisKL20}, that takes into account equilibria reached by risk-averse players.

\textit{Strategy sets of routing games:} They are typically exponential in the number of vertices, hence restricting them is a common assumption. The unnatural character of  instances of routing games, which similarly to Pigou's network exhibit arbitrarily large differences in speeds/latencies between paths,  
 has also been advocated by \cite{lu2012worst}, who assume players have at least one strategy that is not more than \( \lambda \) away from the fastest strategy. Restricting the strategy sets to obtain tighter bounds for PoA is also employed in \citep{bilo2017impact} and \citep{Caragiannis11} for load balancing games (i.e., congestion games where the strategies of players are singleton sets) and for symmetric network congestion games \citep{CorreaJKU19}.  \cite{fotakis2010congestion} proved a pure PoA bound for 
 symmetric atomic congestion games on extension-parallel networks, an interesting class of networks with linearly independent paths, that is equal to that of non-atomic congestion games.
  
\textit{Primal-dual techniques for bounding the Price of Anarchy:} In non-cooperative games, such techniques have been proposed by \cite{B18},  \cite{KM15}, \cite{NR10} and  \cite{T19}. The methods proposed in \citep{B18} and \citep{NR10} operate by explicitly formulating the problem of maximizing the Price of Anarchy of a class of games. Despite using the same formulation, they differ in the choice of the variables. While \cite{NR10} uses the probability distributions defining the outcomes occurring in the formulation, \cite{B18} adopts suitable multipliers for the resource cost functions. The methods in \citep{KM15} and \citep{T19}, instead, build on a formulation for the problem of optimizing the social function, and then implement the equilibria conditions within the choice of the dual variables. We adopt the method proposed in \citep{B18} as it appears to be more flexible and powerful in our realm of application. The first advantage is that it generalizes to any type of cost functions, while all the others require some restrictions: the method in \citep{NR10} can only be applied to affine functions, the one in \citep{KM15} requires convex functions, while that of \cite{T19} needs non-decreasing ones. Secondly, the method (if properly used) always yields tight bounds on the Price of Anarchy, while those in \citep{KM15} and \citep{T19} are limited by the integrality gap of the formulation. Last but not least, it models in a simple, direct and intuitive way any new twist, as the free-flow property considered in this work, one may want to add to the scenario of application. 

Further generalizations or variants of the above primal-dual techniques have been also considered in the context of algorithmic design, with the aim of improving the Price of Anarchy in non-cooperative games (see, for instance, \cite{BiloV19,BiloV19stack,PaccagnanCFM21,VijayalakshmiS20}). With this respect, \cite{PaccagnanCM20} and \cite{PaccagnanM22} propose a general primal-dual framework to design the players utility/cost functions so as to optimize the Price of Anarchy, and such functions are defined as solutions of some tractable linear programs. 

\textit{Transportation research and estimation of Price of Anarchy:} The seminal work of  \cite{Wardrop52} introduces and formalizes one of the first notions of equilibrium in transportation networks. A proof of the equal social costs for equilibria and optimum (i.e., \( \poa = 1 \)) in parallel links routing games appears in  \citep{nagurney2009relative}. Related ideas from sensitivity analysis for edge cost functions are treated in  \citep{tobin1988sensitivity}. Similarly, deviations from the perfectly rational user equilibrium have been investigated in \cite{Takalloo2020} by quantified the worst-case analytical bounds. Moreover, current technological developments on vehicles' connectivity and automation have created new opportunities to carefully quantify the actual benefits of large-scale transportation systems. Along the way, several methodological techniques have been proposed to study the magnitude of PoA in real traffic networks, ranging from heuristics \citep{Youn2008} to algorithms of nonconvex optimization \citep{zhang2018price} and to microscopic simulation \citep{Belov2021}. For instance, the Price of Anarchy was estimated for the city of Boston with different means from our study by  \cite{zhang2018price}, where the sensitivity of the social cost at equilibrium with respect to edge parameters is also discussed. The previously cited works rely on the BPR estimation of cost functions \citep{bureau1964traffic}, which are included in the family of weakly monomial latency functions we define in Section~\ref{sec:model}. The free-flow property in transportation networks has been first proposed by \cite{J+05} with respect to the problem of optimizing a centralized traffic flow without imposing too longer detours to some users.

\label{sec:Our  Contribution}

\section{Model and Definitions}
\label{sec:model}
For a positive integer $i$, let $[i]:=\{1,2,\ldots,i\}$. Given a set $A$ and a set $B\supseteq A$, let $\chi_A:B\rightarrow \{0,1\}$ denote the indicator function, i.e., $\chi_A(x)=1$ if $x\in A$ and $\chi_A(x)=0$ if $x\notin A$. Given a tuple of numbers $(\alpha_1,\alpha_2,\ldots, \alpha_k)$, we write $(\alpha_1,\alpha_2,\ldots, \alpha_k)>0$ if $\alpha_i\geq 0$ for any $i\in [k]$ and $\alpha_i>0$ for some $i\in [k]$.

\paragraph{\bf Non-atomic Congestion Games. } A {\em non-atomic congestion game} (from now on, simply a {\em congestion game}) is a tuple $\CG=\left(\N,(r_i)_{i\in \N},E,(\ell_e)_{e\in E},(\Sigma_i)_{i\in\N}\right)$, where $[n]$ is a set of types, $E$ is a set of resources, $\ell_e:\RPP\rightarrow\RPP$ is the {\em latency function} of resource $e\in E$, and, for each $i\in\N$, $r_i\in\RP$ is the {\em amount of players} of type $i$ and $\Sigma_i\subseteq 2^E\setminus \emptyset$ is the {\em set of strategies} for players of type $i$ (i.e. a strategy is a non-empty subset of resources). We assume that latency functions are non-decreasing, positive, and
 continuous\endnote{The property of continuity is well-motivated by most of the real-life scenarios modelled by non-atomic congestion games. Anyway, our theoretical results hold even with the weaker assumption of right-continuity.}.

\paragraph{\bf Classes of Congestion Games.} A {\em network congestion game} is a congestion game based on a graph $G=(V,E)$, where the set of resources coincides with $E$, each type $i$ is associated with a pair of nodes $(u_i,v_i)\in V\times V$, so that the set of strategies of players of type $i$ is the set of paths from $u_i$ to $v_i$ in graph $G$. If there exists $u^*\in V$ such that $u^*=u_i$ for any $i\in \N$, the game is called {\em single-source network congestion game}. Let $\mathcal{P}$ be the set of all the paths $P$ connecting source $u_i$ with destination $v_i$, for any pair source-destination $(u_i,v_i)$. The game is called {\em path-disjoint network congestion game} if all the paths in $\mathcal{P}$ are pair-wise node-disjoint. 

A {\em load balancing game} is a congestion game in which each strategy is a singleton, i.e., $S=\{e\}$ for some $e\in E$, for any strategy $S\in \Sigma_i$ and type $i\in [n]$. A {\em parallel-link game} (or {\em symmetric load balancing game}) is a load balancing game in which all players have the same set of strategies. It is well-known that each load balancing game (resp. parallel-link game) can be modelled as a single-source congestion game (resp. path-disjoint network congestion game).

\paragraph{\bf Latency Functions.} For the sake of simplicity, we extend the domain of each latency function $\ell(x)$ to $x=0$ in such a way that $\ell(0)=\lim_{x\rightarrow 0^+}\ell(x)$. Given a class of latency functions $\mathcal{F}$, let $[\mathcal{F}]_H:=\{f:f(x)=g(x)-g(0),\ g\in\mathcal{F}\}$. Observe that $f(0)=0$ for any $f\in[\mathcal{F}]_H$ by definition. In the following, we use similar definitions as in \citep{R03}. $\mathcal{F}$ is {\em homogeneous} if $[\mathcal{F}]_H=\mathcal{F}$. $\mathcal{F}$ is {\em weakly diverse} if $[\mathcal{F}]_H\subseteq\mathcal{F}$ and there exists a constant latency function in $\mathcal{F}$ (i.e., a function $g$ such that $g(x)=	\beta$ for any $x$, for some $\beta>0$). $\mathcal{F}$ is {\em scale-closed} if it contains all the functions $f$ such that $f(x)=\alpha g(x)$, for any $g\in\mathcal{F}$ and $\alpha>0$. $\mathcal{F}$ is {\em strongly diverse} if contains all the functions $f$ such that $f(x)=\alpha g(x)+\beta$, for any $g\in[\mathcal{F}]_H$ and $(\alpha,\beta)>0$; observe that, if $\mathcal{F}$ is strongly diverse, then it is both weakly diverse and scale-closed.

A {\em polynomial latency function} of maximum degree $p$ and minimum degree $q$ (with $p\geq q\geq 1$) is defined as $\ell_e(x):=\sum_{d=q}^p\alpha_{e,d}x^d+\beta_e$, where $(\alpha_{e,q},\alpha_{e,q+1},\ldots, \alpha_{e,p},\beta_e)>0$. Let $\mathcal{P}_{p,q}$ denote the class of polynomial latency functions of maximum degree $p$ and minimum degree $q$; observe that, if $1\leq q'\leq q\leq p\leq p'$, then $\mathcal{P}_{p,q}\subseteq \mathcal{P}_{p',q'}$. A {\em weakly monomial latency function} of degree $p$ is defined as $\ell_e(x):=\alpha_{e,p}x^p+\beta_e$, with $(\alpha_e,\beta_e)>0$. In the previous definition, $\ell_e$ is called {\em monomial latency function} of degree $p$ if $\beta_e=0$. Let $\mathcal{WM}_p$ (resp. $\mathcal{M}_p$) denote the class of weakly monomial latency functions (resp. monomial latency functions). Observe that $\mathcal{M}_p\subset \mathcal{WM}_p=\mathcal{P}_{p,p}$ for any integer $p\geq 1$. A latency function $\ell_e$ is {\em affine} if $\ell_e\in \mathcal{WM}_1$, and it is {\em linear} if $\ell_e\in \mathcal{M}_1$.

\paragraph{\bf Strategy Profiles and Pure Nash Equilibria.} A {\em strategy profile} is a tuple $\sg:=(\sigma_{i,S})_{i\in \N,S\in \Sigma_i}$ with $\sum_{S\in \Sigma_i}\sigma_{i,S}=r_i$ for any $i\in \N$, that is a state of the game where $\sigma_{i,S}\geq 0$ is the total amount of players of type $i$ selecting strategy $S$ for any $i\in \N$ and $S\in \Sigma_i$. Given a strategy profile $\sg$, $k_e(\sg):=\sum_{i\in \N,S\in \Sigma_i:e\in S}\sigma_{i,S}$ is the {\em congestion} of $e$ in $\sg$, i.e., the total amount of players selecting $e$ in $\sg$, and given a strategy $S$, $c_{S}(\sg):=\sum_{e\in S}\ell_e(k_e(\sg))$ is the {\em cost} of players selecting $S$ in $\sg$. A strategy profile $\sg$ is a {\em pure Nash equilibrium} (or {\em Wardrop equilibrium}, or {\em equilibrium flow}) if and only if, for each $i\in\N$, $S\in\Sigma_i:\sigma_{i,S}>0$ and $S'\in\Sigma_i$, it holds that $c_{S}(\sg)\leq c_{S'}(\sg)$.

\paragraph{\bf Quality of Equilibria.}
A {\em social function} that is usually used as a measure of the quality of a strategy profile in  congestion games is the {\em total latency}, defined as ${\sf SUM}({\bm \sigma}):=\sum_{e\in E}k_e(\sg)\ell_e(k_e(\sg))=\sum_{i\in\N}r_i c_i({\bm \sigma})$ at equilibrium ${\bm \sigma}$. A {\em social optimum} is a strategy profile ${\bm \sigma}^*$ minimizing $\sf SUM$.

The {\em Price of Anarchy} of a congestion game $\sf CG$ (with respect to the social function $\sf SUM$), denoted as $\sf {\sf PoA}(\CG)$, is the supremum of the ratio ${\sf SUM}({\bm\sigma})/{\sf SUM}({\bm\sigma}^*)$, where $\bm\sigma$ is a pure Nash equilibrium for $\sf CG$ and ${\bm\sigma}^*$ is a social optimum for $\CG$. As shown in \cite{roughgarden2002bad}, all pure Nash equilibria of any congestion game have the same total latency. Thus, the Price of Anarchy can be redefined as the ratio ${\sf SUM}({\bm\sigma})/{\sf SUM}({\bm\sigma}^*)$, where $\bm\sigma$ is an arbitrary pure Nash equilibrium for $\sf CG$ and ${\bm\sigma}^*$ is a social optimum for $\CG$.

\paragraph{\bf Free-Flow Congestion Games}
Given $\theta\in [0,\infty]$, a $\theta$-free-flow congestion game $\CG_\theta$ is a congestion game in which, for each $i\in\N$ and $S,S'\in\Sigma_i$, it holds that $\sum_{e\in S}\ell_e(0)\leq (1+\theta)\sum_{e\in S'}\ell_e(0)$, i.e., all the strategies available to players of type $i$, when evaluated in absence of congestion, are within a factor $1+\theta$ one from the other. Observe that free-flow congestion games are congestion games obeying some special properties. Thus, all positive results holding for congestion games carries over to $\theta$-free-flow congestion games for any value of $\theta$. Moreover, for $\theta=\infty$, any congestion game is a $\theta$-free-flow congestion game.

\begin{example}
A simple evidence of how the value of $\theta$ affects the Price of Anarchy of $\theta$-free-flow congestion games, is given by a parallel-link game $\CG_c$ with a unitary amount of players, and two resources $e_1,e_2$ having latency functions defined as $\ell_1(x)=1$ and $\ell_2(x)=x+c$, for some $c\in [0,1]$. We observe that, for $c=0$, we get the usual Pigou network that matches the worst-case Price of Anarchy of $4/3$ achieved by classical non-atomic congestion games \citep{roughgarden2002bad}; however, this network constitutes an extreme case of $\theta$-free-flow congestion games with $\theta=\infty$, since $1+\theta=\ell_1(0)/\ell_2(0)=1/0=\infty$. 

Conversely, if $c>0$, we have that $\CG_c$ is a $\theta$-free-flow congestion game with $\theta=1/c-1$ (since $1+\theta=\ell_1(0)/\ell_2(0)=1/c$). Furthermore, the Price of Anarchy becomes equal to $4/((c+1)(3-c))<4/3\approx 1.333$, i.e., the social performance increases as $\theta=1/c-1$ decreases. Indeed, the unique equilibrium flow $\bm\sigma$ is obtained when all players select resource $e_1$, and the social optimum $\bm\sigma^*$ is obtained when $(1+c)/2$ players select resource $e_1$, and the remaining ones select $e_2$; thus, by simple calculations, we get $$\poa(\CG)=\frac{{\sf SUM}({\bm\sigma})}{{\sf SUM}({\bm\sigma}^*)}=\frac{1}{(c+1)(3-c)/4}=\frac{4}{(c+1)(3-c)}.$$ We observe that, even for $\theta=1$ (attained by $c=1/2$), we get a Price of Anarchy of $16/15\approx 1.066$, thus showing that the equilibrium-flows of the considered congestion game guarantee a social performance that is very close to the optimal one, if the value of $\theta$ is sufficiently small. 
\end{example}

\section{Price of Anarchy of Free-Flow Congestion Games}
In this section, we give tight bounds on the Price of Anarchy of free-flow congestion games. Before going into details, we sketch the high level building blocks of the proofs of the upper bounds. For general $\theta$-free-flow congestion games, by adapting the primal-dual method \citep{B18}, we formulate the problem of bounding the Price of Anarchy by means of a factor-revealing pair of primal-dual linear programs. The techniques work as follows.

Given a $\theta$-free-flow congestion game $\CG_\theta$ and a family of latency functions $\mathcal{F}$, we know that we can model the latency of every resource $e\in E$ as $\ell_e(x)=\alpha_e f_e(x)+\beta_e$, with $f_e\in [\mathcal{F}]_H$, $\alpha_e\in\{0,1\}$ and $\beta_e\geq 0$. We fix a Nash equilibrium $\sg$ and a social optimum $\sg^*$ for $\CG_\theta$. Hence, for every $e\in E$, the congestions $k_e(\sg)$ and $k_e(\sg^*)$ of $e$ in $\sg$ and $\sg^*$, respectively, become fixed constants. As the Price of Anarchy measures the worst-case ratio of ${\sf SUM}(\sg)$ over ${\sf SUM}(\sg^*)$, our goal becomes that of choosing suitable values for $\alpha_e$ and $\beta_e$, for every $e\in E$, so as to maximize ${\sf SUM}(\sg)$ under the assumption that ${\sf SUM}(\sg^*)=1$, $\sg$ is a Nash equilibrium and $\CG_\theta$ is a $\theta$-free-flow game. In particular, constraint ${\sf SUM}(\sg^*)=1$ can be assumed without loss of generality by a simple scaling argument, provided we relax the condition $\alpha_e\in\{0,1\}$ with $\alpha_e\geq 0$. Thus, an optimal solution to the resulting linear program, call it {\sf LP}, provides an upper bound to the Price of Anarchy of $\CG_\theta$.

Next step is to compute and analyze the dual of {\sf LP}, that we call {\sf DLP}. {\sf DLP} has three variables, namely $x$, $y$ and $\gamma$, with $x\geq 0$, $y\geq 0$ and $\gamma$ defining its objective value. Thus, by the Weak Duality Theorem, any feasible solution $(x,y,\gamma)$ for {\sf DLP} yields an upper bound of $\gamma$ to the optimal solution of {\sf LP} and so an upper bound to the Price of Anarchy of $\CG_\theta$. For each resource $e\in E$, {\sf DLP} has two constraints, namely $c_1(f_e,k_e(\sg),k_e(\sg^*),x,\gamma)$ and $c_2(f_e,k_e(\sg),k_e(\sg^*),y,\gamma)$, respectively associated to the primal variables $\alpha_e$ and $\beta_e$, and providing lower bounds on $\gamma$. In particular, we consider a feasible solution $(x^*,y^*,\gamma^*)$ that is optimal for {\sf DLP}, under the  further constraints $y=\frac{x-1}{1+\theta}$ and $x\geq 1$. 
%
Then, by exploiting the dual constraints, we obtain two significant lower bounds for $\gamma^*$; both bounds depend on the structural properties of the latency functions $f_e$; moreover, the first bound is also influenced by the choice of the optimal dual variable $x$ (i.e., $x^*$), while the second exhibits a dependence from $\theta$. For any class of latency functions $\mathcal{G}$, by using $k$ and $l$ as a shorthand for $k_e(\sg)$ and $k_e(\sg^*)$, respectively, these lower bounds bounds for $\gamma^*$ are at most equal to $\gamma([\mathcal{F}]_H)$ and $\gamma_\theta([\mathcal{F}]_H)$, with
\begin{align}
&\gamma(\mathcal{G}):=\inf_{x>1}\sup_{k> 0,l>0,f\in \mathcal{G}}\left(\frac{k+x(-k+l)}{l}\right)\frac{f(k)}{f(l)},\label{def_poa_1}\\
&\gamma_\theta(\mathcal{G}):=\sup_{k>l>0,f\in \mathcal{G}}\frac{(k-l)f(k)+kf(k)\theta}{(k-l)f(k)+[(k-l)f(k)+lf(l)]\theta};\label{def_poa_2}
\end{align}
thus the maximum value between $\gamma([\mathcal{F}]_H)$ and $\gamma_\theta([\mathcal{F}]_H)$ is an upper bound on the optimal solution of {\sf DLP}, and then an upper bound on the Price of Anarchy (see Theorem \ref{thm_gen_upp}).
\begin{remark}\label{rema1}
Given a class of latency functions $\mathcal{G}$, we observe that $\gamma(\mathcal{G})$ is at least equal to 1; to show this, it is sufficient setting $k=l$ in the quantities of which the supremum is taken (when defining $\gamma(\mathcal{G})$). Furthermore, we observe that $\gamma_\theta(\mathcal{G})=1$  for $\theta=0$ and that $\gamma_\theta(\mathcal{G})$ is non-decreasing in $\theta\geq 0$. 
\end{remark}
An important advantage of the primal-dual method is that, whenever {\sf LP} provides a tight characterization of the properties possessed by the games and the equilibria under analysis, an optimal solution to {\sf DLP} can be fruitfully exploited to construct, quite systematically, but not without effort, matching lower bounding instances. We manage to achieve this result also in this case, but, given the very technical nature of the constructions, we refer the interested reader to the appendix.

In the related literature, bounds on the Price of Anarchy are often obtained by exploiting Roughgarden's smoothness framework \citep{roughgarden2015intrinsic}\endnote{Other variants of the smoothness framework are introduced and applied in \citep{Bach2014,Chan2019,Roug15b,Roug2017}.}. It is based on an inequality linking together the social value of an optimal solution and the sum of the players' costs at an equilibrium, thus requiring the use of two variables. However, for certain settings, as the one considered in this work, additional structural properties of the game need to be embedded in the model. This requires more sophisticated constraints involving a higher number of variables. The primal-dual method handles these twists more easily, as it suffices writing down properly all the additional constraints that need to be satisfied by the model (in our case, the free-flow property). Then, the final set of factor-revealing inequalities that needs to be analyzed elegantly results as a consequence of the duality theory.

For the case of parallel-link and path-disjoint games, we apply a similar, although more direct approach. We fix once again $\CG_\theta$, the family of latency functions $\mathcal{F}$, the latency of every resource $e\in E$, a Nash equilibrium $\sg$ and a social optimum $\sg^*$ for $\CG_\theta$, so as to obtain constant values for both $k_e(\sg)$ and $k_e(\sg^*)$. This time, instead of resorting to linear programming, we write down the parametric expression of the Price of Anarchy as a function of $k_e(\sg)$, $k_e(\sg^*)$ and the latency functions of the resources in the game. A key feature of this case, that makes it different from the general setting analyzed before, is that, here, we need have $\sum_{e\in E}k_e(\sg)=\sum_{e\in E}k_e(\sg^*)$. By exploiting this equality, together with the equilibrium conditions and the $\theta$-free-flow property of $\CG_\theta$, we create a sequence of more and more relaxed upper bounds for the Price of Anarchy, until we end up to a sufficiently simple formula. In particular, let
\begin{equation}\label{def_poa_3}
\eta_\theta(\mathcal{G}):=\sup_{k>l>0,f\in\mathcal{G}}\frac{kf(k)+kf(k)\theta}{kf(k)+[(k-l)f(k)+lf(l)]\theta};
\end{equation}
we show that the maximum value between $\gamma([\mathcal{F}]_H)$ and $\eta_\theta([\mathcal{F}]_H)$ is an upper bound on the Price of Anarchy (see Theorem \ref{thm_gen_upp_par}). 
\begin{remark}\label{rema2}
Given a class of latency functions $\mathcal{G}$, we observe that $\eta_0(\mathcal{G})=1$ and that $\eta_\theta(\mathcal{G})$ is non-decreasing in $\theta\geq 0$. Furthermore, we observe that $\eta_\theta(\mathcal{G})\leq \gamma_\theta(\mathcal{G})$ for any $\theta\geq 0$. 
\end{remark}
Also in this case, we can show that the performed analysis is tight by providing matching lower bounding instances whose description is again deferred to the appendix.

For the sake of simplicity, the case $\theta=\infty$ (i.e., congestion games without the free-flow hypothesis) is not considered in the main theorems, as it has been already treated by some previous works (\cite{R03,C+04}) under similar hypothesis on the latency functions. Anyway, we separately treat the case $\theta=\infty$ in a corollary of the main theorems (see Corollary \ref{cor1}), to provide better bounds on the Price of Anarchy (with respect to the existing ones) under weaker hypothesis on the considered classes of latency functions (see Subsection \ref{sub_int_1} for further details).
\subsection{The Main Theorems}
\begin{theorem}\label{thm_gen_upp}
Let $\CG_\theta$ be a $\theta$-free-flow congestion game with latency functions in $\mathcal{F}$ and $\theta\geq 0$. Then $\poa(\CG_\theta)\leq \max\{{\gamma}([\mathcal{F}]_H),{\gamma}_\theta([\mathcal{F}]_H)\}$. Furthermore, this bound is tight for single-source network games if $\mathcal{F}$ is weakly diverse, and even for (non-symmetric) load balancing games if $\mathcal{F}$ is strongly diverse.
\end{theorem}
\proof{Proof:}
Let $\sg=(\sigma_{i,S})_{i\in \N, S\in \Sigma_i}$ and $\sg^*=(\sigma_{i,S}^*)_{i\in \N, S\in \Sigma_i}$ be a pure Nash equilibrium and a social optimum for $\CG_\theta$, respectively. Let $k_e:=k_e(\sg)$ and $l_e:=k_e(\sg^*)$ for any $e\in E$. Let $\ell_e(x):=\alpha_e f_e(x)+\beta_e$ be the latency function of each resource $e\in E$, with $\alpha_e\in \{0,1\}$, $\beta_e\geq 0$, and $f_e\in[\mathcal{F}]_H$. By applying the primal-dual method \citep{B18}, we have that the optimal solution of the following linear program in variables $(\alpha_e)_{e\in E}$ and $(\beta_e)_{e\in E}$ is an upper bound on $\poa(\CG_\theta)$:
\begin{align}
{\sf LP:}\ \max \ & \sum_{e\in E}(\alpha_e k_ef_e(k_e)+\beta_e k_e)\nonumber\\
\text{s.t.}\ & \sum_{e\in E}(\alpha_e k_ef_e(k_e)+\beta_e k_e)\leq \sum_{e\in E}(\alpha_e l_ef_e(k_e)+\beta_e l_e)\label{primcost1}\\
& \sum_{e\in E}\beta_el_e\leq \sum_{e\in E}(1+\theta)\beta_ek_e\label{primcost2}\\
&\sum_{e\in E}(\alpha_e l_ef_e(l_e)+\beta_e l_e)=1\label{primcost3}\\
&\alpha_e,\beta_e\geq 0\quad \forall e\in E.\nonumber
\end{align}
Indeed:
\begin{description}
\item[$\bullet$ Objective function:] Each latency function $\ell_e$ can be expressed as $\ell_e(k_e)=\alpha_e f_e(k_e)+\beta_e$ with $\alpha_e\in \{0,1\}$.
\item[$\bullet$ Constraint (\ref{primcost1}):] For any $i\in \N$, and any two strategies $S,S^*\in \Sigma_i$, let $\sigma_{i,S,S^*}\geq 0$ denote the amount of players of type $i$ selecting $S$ in $\sg$ and selecting $S^*$ in $\sg^*$. By the pure Nash equilibrium conditions, we have that $\sum_{e\in S}(\alpha_e f_e(k_e)+\beta_e)\leq \sum_{e\in S^*}(\alpha_e f_e(k_e)+\beta_e)$, for any $i\in [n]$, and for any two strategies $S, S^*\in \Sigma_i$ such that $\sigma_{i,S,S^*}>0$. Then, we have that
\begin{align}
0&\geq \sum_{i\in \N}\sum_{S,S^*\in \Sigma_i}\sigma_{i,S,S^*}\left(\sum_{e\in S}(\alpha_e f_e(k_e)+\beta_e)- \sum_{e\in S^*}(\alpha_e f_e(k_e)+\beta_e)\right)\nonumber\\
&=\sum_{e\in E}\left(\sum_{i\in \N,S,S^*\in \Sigma_i:e\in S}\sigma_{i,S,S^*}\right)(\alpha_e f_e(k_e)+\beta_e)- \sum_{e\in E}\left(\sum_{i\in \N,S,S^*\in \Sigma_i:e\in S^*}\sigma_{i,S,S^*}\right)(\alpha_e f_e(k_e)+\beta_e)\nonumber\\
&=\sum_{e\in E}\left(\sum_{i\in \N,S\in \Sigma_i:e\in S}\sum_{S^*\in\Sigma_i}\sigma_{i,S,S^*}\right)(\alpha_e f_e(k_e)+\beta_e)- \sum_{e\in E}\left(\sum_{i\in \N,S^*\in \Sigma_i:e\in S^*}\sum_{S\in\Sigma_i}\sigma_{i,S,S^*}\right)(\alpha_e f_e(k_e)+\beta_e)\nonumber\\
&=\sum_{e\in E}\left(\sum_{i\in \N,S\in \Sigma_i:e\in S}\sigma_{i,S}\right)(\alpha_e f_e(k_e)+\beta_e)- \sum_{e\in E}\left(\sum_{i\in \N,S^*\in \Sigma_i:e\in S^*}\sigma_{i,S^*}\right)(\alpha_e f_e(k_e)+\beta_e)\nonumber\\
&=\sum_{e\in E}(\alpha_e k_ef_e(k_e)+\beta_e k_e)-\sum_{e\in E}(\alpha_e l_ef_e(k_e)+\beta_e l_e),\nonumber
\end{align}
and this implies constraint (\ref{primcost1}).
\item[$\bullet$ Constraint (\ref{primcost2}):] By using the definition of $\theta$-free-flow congestion games, we have that $\sum_{e\in S^*}\beta_e \leq (1+\theta) \sum_{e\in S}\beta_e$, for any $i\in [n]$, and for any strategies $S, S^*\in \Sigma_i$ such that $\sigma_{i,S,S^*}>0$. Thus
\begin{align}
0&\geq \sum_{i\in \N}\sum_{S,S^*\in \Sigma_i}\sigma_{i,S,S^*}\left(\sum_{e\in S^*}\beta_e- (1+\theta) \sum_{e\in S}\beta_e\right)\nonumber\\
&=\sum_{e\in E}\left(\sum_{i\in \N,S,S^*\in \Sigma_i:e\in S^*}\sigma_{i,S,S^*}\right)\beta_e- (1+\theta) \sum_{e\in E}\left(\sum_{i\in \N,S,S^*\in \Sigma_i:e\in S}\sigma_{i,S,S^*}\right)\beta_e\nonumber\\
&=\sum_{e\in E}\left(\sum_{i\in \N,S^*\in \Sigma_i:e\in S^*}\sigma_{i,S^*}\right)\beta_e- (1+\theta) \sum_{e\in E}\left(\sum_{i\in \N,S\in \Sigma_i:e\in S}\sigma_{i,S}\right)\beta_e\nonumber\\
&=\sum_{e\in E}l_e\beta_e-\sum_{e\in E}(1+\theta) k_e\beta_e,\nonumber
\end{align}
and this implies constraint (\ref{primcost2}).
\item[$\bullet$ Constraint (\ref{primcost3}) + Relaxations:] The Price of Anarchy of $\CG_\theta$ is $\frac{{\sf SUM}(\sg)}{{\sf SUM}(\sg^*)}= \frac{\sum_{e\in E}(\alpha_e k_ef_e(k_e)+\beta_e k_e)}{\sum_{e\in E}(\alpha_e l_ef_e(l_e)+\beta_e l_e)}$, thus, by maximizing such value over all the possible $\alpha_e,\beta_e\geq 0$ subject to constraints (\ref{primcost1}) and (\ref{primcost2}), we get an upper bound on the Price of Anarchy of $\CG_\theta$. If we restrict the values of $\alpha_e$ and $\beta_e$ in such a way that the further normalization constraint (\ref{primcost3}) holds, we do not affect the maximum value considered above, thus finding such maximum value is equivalent to find the optimal solution of {\sf LP}. This can be achieved by relaxing the condition $\alpha_e\in\{0,1\}$ to $\alpha_e\geq 0$.
\end{description}

We call {\em generating set} a generic finite set $\mathcal{T}\subseteq \{(k,l,f):k,l\geq 0,f\in[\mathcal{F}]_H\}$. For any generating set $\mathcal{T}$, we define a linear program in variables $\gamma,x,y$:
\begin{align}
{\sf DLP}(\mathcal{T})\quad  \min \quad & \gamma\nonumber\\
\text{s.t.} \quad & lf(l)\gamma\geq kf(k)+x(-kf(k)+lf(k))\quad \forall (k,l,f)\in \mathcal{T}\label{dualcost1}\\
& l\gamma\geq k+x(-k+l)+y(-l+(1+\theta) k)\quad \forall (k,l,f)\in \mathcal{T}\label{dualcost2}\\
& x,y\geq 0,\nonumber
\end{align}
Let $\mathcal{T}_E:=\{(k_e,l_e,f_e):e\in E\}$. Observe that ${\sf DLP}(\mathcal{T}_E)$ is the dual of {\sf LP}. Indeed, for $\mathcal{T}=\mathcal{T}_E$, each dual constraint of type (\ref{dualcost1}) (resp. (\ref{dualcost2})) is associated to some primal variable $\alpha_e$ (resp. $\beta_e$), and $x$ (resp. $y$, resp. $\gamma$) is the dual variable associated to the primal constraint defined in (\ref{primcost1}) (resp. (\ref{primcost2}), resp. (\ref{primcost3})).

By the Weak Duality Theorem, any feasible solution $(x,y,\gamma)$ of ${\sf DLP}(\mathcal{T}_E)$ is such that $\gamma$ is at least equal to the optimal value of {\sf LP}, that is an upper bound on $\poa(\CG_\theta)$. Thus, to show the claim, it is sufficient providing, for any generating set $\mathcal{T}$, a tuple $(x,y,\gamma)$ with $\gamma\leq\max\{{\gamma}([\mathcal{F}]_H),{\gamma}_\theta([\mathcal{F}]_H)\}$ that is a feasible solution of ${\sf DLP}(\mathcal{T})$. 

Let $\mathcal{T}$ be an arbitrary generating set. Let $(x^*,y^*,\gamma^*)$ be the optimal solution of ${\sf DLP}(\mathcal{T})$, subject to the further constraints $x\geq 1$ and $y:=\frac{x-1}{1+\theta}$. Under the above (further) constraints on ${\sf DLP}(\mathcal{T})$, we have that the constraints of type \eqref{dualcost1} and \eqref{dualcost2} of ${\sf DLP}(\mathcal{T})$ are satisfied by any value of $\gamma$ if $l=0$, thus we can assume without loss of generality that $l$ is always positive. We conclude that $(x^*,\gamma^*)$ is the optimal solution of the following linear program:
\begin{align}
{\sf DLP2}(\mathcal{T}):\quad \min \ & \gamma\nonumber\\
\text{s.t.} \ & \gamma\geq \frac{kf(k)+x(-kf(k)+lf(k))}{lf(l)},\ \forall (k,l,f)\in\mathcal{T}: l>0\label{dualcost12}\\
& \gamma\geq \frac{x\theta+1}{1+\theta}\label{dualcost22}\\
& x\geq 1,\label{dualcost32}
\end{align}
where \eqref{dualcost22} is obtained by imposing $y:=\frac{x-1}{1+\theta}$ and $l>0$ in \eqref{dualcost2}.

First of all, we assume that $(x^*,\gamma^*)$ is such that constraint \eqref{dualcost22} is not tight. In such case, we have that $(x^*,\gamma^*)$ continues to be the optimal solution of a linear program obtained from ${\sf DLP2}(\mathcal{T})$ by deleting constraint \eqref{dualcost22}. Thus, we get\endnote{Equality \eqref{upp_form_1_prel} holds since, by continuity, the supremum over $k>0$ is equal to that over $k\geq 0$.}
\begin{align}
\poa(\CG_\theta) &\leq\gamma^*\label{upp_form_1.000}\\
&=\max_{(k,l,f)\in \mathcal{T}: l> 0 }\frac{kf(k)+x^*(-kf(k)+lf(k))}{lf(l)}\nonumber\\
&=\min_{x\geq 1} \max_{(k,l,f)\in \mathcal{T}: l> 0 }\frac{kf(k)+x(-kf(k)+lf(k))}{lf(l)}\nonumber\\
&\leq \inf_{x\geq 1}\sup_{k\geq 0, l>0, f\in [\mathcal{F}]_H}\frac{kf(k)+x(-kf(k)+lf(k))}{lf(l)}\nonumber\\
&=\inf_{x\geq 1}\sup_{k> 0, l>0, f\in [\mathcal{F}]_H}\frac{kf(k)+x(-kf(k)+lf(k))}{lf(l)}\label{upp_form_1_prel}\\
&\leq \inf_{x> 1}\sup_{k> 0, l>0, f\in [\mathcal{F}]_H}\frac{kf(k)+x(-kf(k)+lf(k))}{lf(l)}\nonumber\\
&= {\gamma}([\mathcal{F}]_H),\label{upp_form_1}
\end{align}
and the claim follows when constraint \eqref{dualcost22} is not tight under solution $(x^*,\gamma^*)$. 

Now, we assume that $(x^*,\gamma^*)$ is such that constraint \eqref{dualcost22} is tight, i.e., $\gamma^*=\frac{x^*\theta+1}{1+\theta}$. As ${\sf DLP2}(\mathcal{T})$ has three variables, we have that the optimal solution $(x^*,\gamma^*)$ can be chosen in such a way that a further constraint is tight. If such tight constraint is \eqref{dualcost32}, i.e., $x^*=1$ holds, by exploiting constraint~\eqref{dualcost22}  we get $\gamma^*=1$; thus $\poa(\CG_\theta)\leq \gamma^*=1\leq {\gamma}_\theta([\mathcal{F}]_H)$, and the claim follows. If the further tight constraint is \eqref{dualcost12} for some triple $(k,l,f)\in \mathcal{T}$, we have that pair $(x^*,\gamma^*)$ satisfies equations $\gamma^*=\frac{kf(k)+x^*(-kf(k)+lf(k))}{lf(l)}$ and $\gamma^*=\frac{x^*\theta+1}{1+\theta}$, that is, $\gamma^*=\frac{(k-l)f(k)+kf(k)\theta}{(k-l)f(k)+[(k-l)f(k)+lf(l)]\theta}$. Thus, we get
\begin{align}
\poa(\CG_\theta) &\leq\gamma^*\nonumber\\
&=\frac{(k-l)f(k)+kf(k)\theta}{(k-l)f(k)+[(k-l)f(k)+lf(l)]\theta}\nonumber\\
&\leq \sup_{k>l>0,f\in [\mathcal{F}]_H}\frac{(k-l)f(k)+kf(k)\theta}{(k-l)f(k)+[(k-l)f(k)+lf(l)]\theta}\nonumber\\
&= {\gamma}_\theta([\mathcal{F}]_H),\label{upp_form_2}
\end{align}
and this concludes the proof for the upper bound. 

The construction of the matching lower bounding instances is deferred to the appendix.
\Halmos 

By using the same proof arguments of the above theorem, we get the following corollary (whose proof is deferred to the appendix), that provides tight bounds on the Price of Anarchy for classical congestion games (i.e., without the $\theta$-free-flow hypothesis). 
\begin{corollary}\label{cor1}
The Price of Anarchy of congestion games with latency functions in $\mathcal{F}$ is at most $\gamma(\mathcal{F})$. Furthermore, this bound is tight for parallel-link games if $\mathcal{F}$ is scale-closed, and is tight for path-disjoint network-congestion games if $\mathcal{F}$ is arbitrary.
\end{corollary}

We now show that, when considering either parallel-link games or path-disjoint network congestion games, a better bound on the Price of Anarchy can be achieved. 
\begin{theorem}\label{thm_gen_upp_par}
Let $\PLG_\theta$ be a $\theta$-free-flow path-disjoint network congestion game with latency functions in $\mathcal{F}$ and $\theta\geq 0$. Then $\poa(\PLG_\theta)\leq \max\{{\gamma}([\mathcal{F}]_H),{\eta}_\theta([\mathcal{F}]_H)\}$. Furthermore, this bound is tight if $\mathcal{F}$ is weakly diverse, and even for parallel-link games if $\mathcal{F}$ is strongly diverse.
\end{theorem}
\proof{Proof:}
Here, we show the claim for the restricted case of $\theta$-free-flow parallel-link games only. Indeed, the case of path-disjoint network games (which is formally treated in the appendix) can be reduced to that of parallel-link games, after replacing each path $P$ of the input instance with a single resource $e_P$ with latency defined as $\overline{\ell}_{e_P}(x):=\sum_{e\in P}\ell_e(x)$. 

Let $\PLG_\theta$ be a $\theta$-free-flow parallel-link game with latency functions in $\mathcal{F}$. Let $\ell_e(x):=\alpha_e f_e(x)+\beta_e$ be the latency function of each resource $e\in E$, with $\alpha_e=\{0,1\}$, $\beta_e\geq 0$, and $f_e\in [\mathcal{F}]_H$. Let $\sg$ and $\sg^*$ be a pure Nash equilibrium and a social optimum for $\PLG_\theta$, respectively. Let $k_e:=k_e(\sg)$ and $l_e:=k_e(\sg^*)$ for any $e\in E$. Let $E^+:=\{e\in E:k_e>l_e\}$ and $E^-:=\{e\in E:k_e<l_e\}$. If one set among $E^+$ and $E^-$ is empty, then $E^+=E^-=\emptyset$ necessarily, and we have that the Price of Anarchy of $\PLG_\theta$ is $1$. Then, as ${\eta}_\theta(\mathcal{F})\geq 1$, the claim holds. Thus, we assume that $E^+\neq \emptyset$ and then $E^-\neq \emptyset$. Furthermore, we assume without loss of generality that there are no resources $e\in E$ such that $k_e=l_e$, otherwise, by removing these resources and their users from the game, the Price of Anarchy does not decrease. 

We observe that, we can define a quantity $w_{u,v}>0$ for any two resources $u\in E^+$ and $v\in E^-$ such that, starting from $\sg$, if we shift an amount of flow $w_{u,v}$ from any resource $u\in E^+$ to any resource $v\in E^-$, we get strategy profile $\sg^*$; we observe that  $k_u-l_u=\sum_{v\in E^-}w_{u,v}$ and $l_v-k_v=\sum_{u\in E^+}w_{u,v}$. For any two resources $u\in E^+$ and $v\in E^-$, let $\xi_{u,v}:=w_{u,v}/(k_u-l_u)$ and $\psi_{u,v}:=w_{u,v}/(l_v-k_v)$. By construction of $\xi_{u,v}$ and $\psi_{u,v}$, we get $\sum_{v\in E^-}\xi_{u,v}=1$ for any $u\in E^+$ and $\sum_{u\in E^+}\psi_{u,v}=1$ for any $v\in E^-$. We have that
\begin{align}
{\sf SUM}(\sg)&=\sum_{u\in E^+}k_u\ell_u(k_u)+\sum_{v\in E^-}k_v\ell_v(k_v)\nonumber\\
&=\sum_{u\in E^+}k_u\ell_u(k_u)\sum_{v\in E^-}\xi_{u,v}+\sum_{v\in E^-}k_v\ell_v(k_v)\sum_{u\in E^+}\psi_{u,v}\nonumber\\
&=\sum_{u\in E^+,v\in E^-}(\xi_{u,v} k_u\ell_u(k_u)+\psi_{u,v}k_v\ell_v(k_v))\nonumber\\
&=\sum_{u\in E^+,v\in E^-}\left(\frac{w_{u,v}}{k_u-l_u}k_u\ell_u(k_u)+\frac{w_{u,v}}{l_v-k_v}k_v\ell_v(k_v)\right)\nonumber\\
&=\sum_{u\in E^+,v\in E^-}w_{u,v}\left(\frac{k_u}{k_u-l_u}\ell_u(k_u)+\frac{k_v}{l_v-k_v}\ell_v(k_v)\right)\label{eq_par}
\end{align}
and
\begin{align}
{\sf SUM}(\sg^*)&=\sum_{u\in E^+}l_u\ell_u(l_u)+\sum_{v\in E^-}l_v\ell_v(l_v)\nonumber\\
&=\sum_{u\in E^+}l_u\ell_u(l_u)\sum_{v\in E^-}\xi_{u,v}+\sum_{v\in E^-}l_u\ell_v(l_v)\sum_{u\in E^+}\psi_{u,v}\nonumber\\
&=\sum_{u\in E^+,v\in E^-}(\xi_{u,v} l_u\ell_u(l_u)+\psi_{u,v}l_v\ell_v(l_v))\nonumber\\
&=\sum_{u\in E^+,v\in E^-}\left(\frac{w_{u,v}}{k_u-l_u}l_u\ell_u(l_u)+\frac{w_{u,v}}{l_v-k_v}l_v\ell_v(l_v)\right)\nonumber\\
&=\sum_{u\in E^+,v\in E^-}w_{u,v}\left(\frac{l_u}{k_u-l_u}\ell_u(l_u)+\frac{l_v}{l_v-k_v}\ell_v(l_v)\right).\label{opt_par}
\end{align}
By exploiting (\ref{eq_par}) and (\ref{opt_par}) we get:
\begin{align}
\frac{{\sf SUM}(\sg)}{{\sf SUM}(\sg^*)}&=\frac{\displaystyle\sum_{u\in E^+,v\in E^-}w_{u,v}\left(\frac{k_u}{k_u-l_u}\ell_u(k_u)+\frac{k_v}{l_v-k_v}\ell_v(k_v)\right)}{\displaystyle\sum_{u\in E^+,v\in E^-}w_{u,v}\left(\frac{l_u}{k_u-l_u}\ell_u(l_u)+\frac{l_v}{l_v-k_v}\ell_v(l_v)\right)}\nonumber\\
&\leq \max_{u\in E^+,v\in E^-}\frac{\frac{k_u}{k_u-l_u}\ell_u(k_u)+\frac{k_v}{l_v-k_v}\ell_v(k_v)}{\frac{l_u}{k_u-l_u}\ell_u(l_u)+\frac{l_v}{l_v-k_v}\ell_v(l_v)}\nonumber\\
&=\max_{u\in E^+,v\in E^-}\frac{\frac{k_u}{k_u-l_u}(\alpha_uf_u(k_u)+\beta_u)+\frac{k_v}{l_v-k_v}(\alpha_v f_v(k_v)+\beta_v)}{\frac{l_u}{k_u-l_u}(\alpha_uf_u(l_u)+\beta_u)+\frac{l_v}{l_v-k_v}(\alpha_vf_v(l_v)+\beta_v)}.\label{prop1_par}
\end{align}
Let $u\in E^+,v\in E^-$ be the resources maximizing (\ref{prop1_par}), so that (\ref{prop1_par}) is at most $$F(\alpha_u,\beta_u,\alpha_v,\beta_v):=\frac{\frac{k_u}{k_u-l_u}(\alpha_uf_u(k_u)+\beta_u)+\frac{k_v}{l_v-k_v}(\alpha_v f_v(k_v)+\beta_v)}{\frac{l_u}{k_u-l_u}(\alpha_uf_u(l_u)+\beta_u)+\frac{l_v}{l_v-k_v}(\alpha_vf_v(l_v)+\beta_v)}.$$
In the following, we show that $\max\{{\gamma}([\mathcal{F}]_H),\eta_\theta([\mathcal{F}]_H)\}$ is an upper bound to $F(\alpha_u,\beta_u,\alpha_v,\beta_v)$. As $\sg$ is an equilibrium, we have that $\alpha_uf_u(k_u)+\beta_u\leq \alpha_vf_v(k_v)+\beta_v$ (equilibrium condition), and since the game is $\theta$-free-flow, we have that $\beta_v\leq (1+\theta)\beta_u$ ($\theta$-free-flow condition). Only if $k_v=0$, it might be the case that $\alpha_uf_u(k_u)+\beta_u<\alpha_vf_v(k_v)+\beta_v$. In such a case, since $k_v<l_v$, one can reduce the values of $\alpha_v$ and $\beta_v$ as much as possible so as the $\theta$-free-flow condition continues to hold, while guaranteeing that $\alpha_uf_u(k_u)+\beta_u= \alpha_vf_v(k_v)+\beta_v$ holds and the value of $F(\alpha_u,\beta_u,\alpha_v,\beta_v)$ does not decrease. Thus, we assume without loss of generality that the equilibrium condition is tight, i.e., $\alpha_uf_u(k_u)+\beta_u= \alpha_vf_v(k_v)+\beta_v$ holds. As $\beta_v\leq (1+\theta)\beta_u$, and since $k_v<l_v$, we have that, by increasing $\beta_v$ and decreasing $\alpha_v$ as much as possible so that the $\theta$-free-flow condition is tight (i.e., $\beta_v= (1+\theta)\beta_u$), the equilibrium condition is satisfied, and the value $\alpha_v f_v(k_v)+\beta_v$ does not change, we get that  $\alpha_vf_v(l_v)+\beta_v$ does not increase, and then $F(\alpha_u,\beta_u,\alpha_v,\beta_v)$ does not decrease. Thus, we can assume without loss of generality that $\beta_u=\beta_v/(1+\theta)$ (obtained by tightening the $\theta$-free-flow condition) and that $\alpha_u=(\alpha_vf_v(k_v)+(1-1/(1+\theta))\beta_v)/f_u(k_u)$ (obtained by tightening the equilibrium condition). By using these values of $\alpha_u$ and $\beta_u$,
we can prove the following result which yields the claim (the proof is deferred to the appendix):
\begin{lemma}\label{lemma-thm-2}
It holds that $F(\alpha_u,\beta_u,\alpha_v,\beta_v)\leq\max\left\{{\gamma}([\mathcal{F}]_H),\eta_\theta([\mathcal{F}]_H)\right\}.$
\end{lemma}

Also in this case, the construction of the matching lower bounding instances is deferred to the appendix.
\Halmos

\subsection{Polynomial Latency Functions}

As consequence of the previous results, we can determine the exact Price of Anarchy of free-flow congestion games with polynomial latency functions. For the general case, we have the following theorem.
\begin{theorem}\label{thm_gen_pol}
Fix a value $\theta\geq 0$. The Price of Anarchy of $\theta$-free-flow congestion game with polynomial latency functions of maximum degree $p$ and minimum degree $q$ is $\max\left\{{\gamma}([\mathcal{P}_{p,q}]_H),\gamma_\theta([\mathcal{P}_{p,q}]_H)\right\}$, with
\begin{align}
&\gamma_\theta([\mathcal{P}_{p,q}]_H)=\sup_{t>1}\frac{t^{p+1}(1+\theta)-t^p}{t^{p+1}(1+\theta)-t^p(1+\theta)+\theta},\textrm{ and }\label{pol1}\\
&{\gamma}([\mathcal{P}_{p,q}]_H)=\frac{p^p\left(\sqrt[p-q]{\left(\frac{(p+1)^{p+1}q^q}{(q+1)^{q+1}p^p}\right)^{p+1}}\right)}{(p+1)^{p+1}\left(\sqrt[p-q]{\frac{(p+1)^{p+1}q^q}{(q+1)^{q+1}p^p}}-1\right)}\chi_{[p-1]}(q)+\chi_{\{p\}}(q).\label{pol2}
\end{align}
Furthermore, such bounds are tight even for non-symmetric load balancing games and single-source network congestion games. 
\end{theorem}

For parallel-link games and free-flow path-disjoint network games, we get the following result.

\begin{theorem}\label{thm_gen_pol_par}
Fix a value $\theta\geq 0$. The Price of Anarchy of both $\theta$-free-flow parallel-link games and $\theta$-free-flow path-disjoint network congestion games with polynomial latency functions of maximum degree $p$ and minimum degree $q$ is $\max\left\{{\gamma}([\mathcal{P}_{p,q}]_H),\eta_\theta([\mathcal{P}_{p,q}]_H)\right\}$, where 
\begin{equation}\label{pol3}
\eta_\theta([\mathcal{P}_{p,q}]_H)=\sup_{t>1}\frac{t^{p+1}(1+\theta)}{t^{p+1}(1+\theta)-t^{p}\theta+\theta},
\end{equation}
and ${\gamma}([\mathcal{P}_{p,q}]_H)$ is defined as in Theorem \ref{thm_gen_pol}.
\end{theorem}
In Table \ref{table1}, we have listed the PoA bounds of $\theta$-free-flow games (both general games and path-disjoint games) with polynomial latency functions of maximum degree $p\in [4]$ and minimum degree $q\in [4]$, and with $\theta\in \{0,1/2,1\}$. 
\section{Interpretation and Discussion of the Results}\label{sec_inte}

In this section, we provide a detailed discussion of the implications of our theoretical results and how they relate to previous work.

\subsection{Congestion Games with Homogeneous Latency Functions}\label{sub_int_1} Consider the case of $\theta=\infty$, i.e., general congestion games without the free-flow hypothesis. From \cite{R03} and \cite{C+04}, we know that $\gamma_\infty(\mathcal{F}):=\sup_{k>l>0,f\in [\mathcal{F}]_H}\frac{kf(k)}{f(k)(k-l)+lf(l)}$ is an upper bound on the Price of Anarchy of congestion games with latency function in $\mathcal{F}$, and such bound is tight, even for path-disjoint (resp. parallel-link) games, if $\mathcal{F}$ is not homogeneous (resp. scale-closed and not homogeneous). 

By Corollary \ref{cor1} we have that the Price of Anarchy of any congestion game with latency functions in $\mathcal{F}$ is equal to ${\gamma}(\mathcal{F})$, and the tight PoA is attained even for path-disjoint (resp. parallel-link) games if $\mathcal{F}$ is arbitrary  (resp. scale-closed). Hence, as a byproduct of our analysis, we provide a tight bound of ${\gamma}([\mathcal{F}]_H)={\gamma}(\mathcal{F})$ for the case of homogeneous functions, holding even for simple network topologies. These findings close an open problem posed by \cite{R03}, in which he asked if there exists a simple parallel-link game matching the worst-case Price of Anarchy of arbitrary classes of homogeneous latency functions.

Since our upper bound ${\gamma}(\mathcal{F})$ is tight (by Corollary \ref{cor1}), we immediately have that ${\gamma}(\mathcal{F})\leq \gamma_\infty(\mathcal{F})$. Moreover, for certain classes of latency functions, e.g., homogeneous polynomial latency functions, our bounds are strictly better. By Corollary~\ref{cor1}, we have that the Price of Anarchy of congestion games with homogeneous polynomial latency functions is ${\gamma}([\mathcal{P}_{p,q}]_H)$, furthermore, from a direct computation, we have that ${\gamma}([\mathcal{P}_{p,q}]_H)<\gamma_\infty([\mathcal{P}_{p,q}]_H)$, where the value of ${\gamma}([\mathcal{P}_{p,q}]_H)$ has been established in Theorem \ref{thm_gen_pol}, and $\gamma_\infty([\mathcal{P}_{p,q}]_H)=\frac{(p+1)\sqrt[p]{p+1}}{(p+1)\sqrt[p]{p+1}-p}$ (as shown in \citep{R03}). 

Hence, the upper bound $\gamma_\infty(\mathcal{F})$ provided by \cite{R03} and \cite{C+04} might be non-tight for homogeneous latency functions; furthermore, we also obtain that the Price of Anarchy for homogeneous latency functions might be strictly lower than the one for non-homogeneous functions. 

\subsection{Free-Flow Games with Polynomial Latency Functions}\label{fre_pol_sub}
The Price of Anarchy of $\theta$-free-flow congestion games with monomial latency functions of degree $p~\geq~1$ is equal to that of $0$-free-flow congestion games (as monomial latency functions are homogeneous); thus, By Theorem \ref{thm_gen_pol}, the Price of Anarchy is equal to 
$$\max\{{\gamma}([\mathcal{M}_p]_H),{\gamma}_0([\mathcal{M}_p]_H)\}={\gamma}([\mathcal{M}_p]_H)={\gamma}([\mathcal{P}_{p,p}]_H)=1$$ (where the first equality holds since ${\gamma}_0([\mathcal{F}]_H)=1$ for any class $\mathcal{F}$, and the last equality comes from \eqref{pol2}), thus reobtaining a well-known result in the literature (see, for instance, \citep{H06}).
For weakly-monomial latency functions of degree $p\geq 1$, instead, by Theorem \ref{thm_gen_pol}, we get a bound of $$\max\{{\gamma}([\mathcal{WM}_p]_H),\gamma_\theta([\mathcal{WM}_p]_H)\}=\gamma_\theta([\mathcal{WM}_p]_H)=\gamma_\theta([\mathcal{P}_{p,p}]_H)$$ (where the first equality holds since, by applying \eqref{pol2}, we get ${\gamma}([\mathcal{WM}_p]_H)={\gamma}([\mathcal{P}_{p,p}]_H)=1$); thus we get the PoA  bound defined in equality (\ref{pol1}) of Theorem \ref{thm_gen_pol}. For the particular case of affine functions, i.e., class $\mathcal{WM}_{1}$, we reobtain the same bounds of \cite{BV20}. However, we give an improved result, as our lower bounds hold even for load balancing and single-source network congestion games (in \citep{BV20}, tight lower bounds are given for general congestion games only).

For $\theta$-free-flow path-disjoint games with weakly-monomial latency functions of degree $p\geq 1$, by Theorem \ref{thm_gen_pol_par}, the Price of Anarchy gets equal to $$\max\{{\gamma}([\mathcal{WM}_p]_H),\eta_\theta([\mathcal{WM}_p]_H)\}=\eta_\theta([\mathcal{WM}_p]_H)=\eta_\theta([\mathcal{P}_{p,p}]_H)=\frac{(1+\theta)(p+1)^\frac{p+1}{p}}{(1+\theta)(p+1)^\frac{p+1}{p}-\theta p}$$ (where the last equality follows from a direct computation of the bound defined in \eqref{pol3}), and it is tight even for parallel-link games with two resources only. In this case, with respect to affine functions, we improve on the upper bounds given in \citep{BV20}.

\subsection{Simpler Upper Bounds for Path-Disjoint Free-Flow Games}
We observe that
\begin{align*}
&\eta_\theta([\mathcal{F}]_H)=\sup_{k>l>0,f\in[\mathcal{F}]_H}\frac{kf(k)+kf(k)\theta}{kf(k)+[(k-l)f(k)+lf(l)]\theta}\leq \sup_{k>l>0,f\in[\mathcal{F}]_H}\frac{kf(k)(1+\theta)}{kf(k)}=1+\theta,
\end{align*}
thus, by Theorem \ref{thm_gen_upp_par}, the Price of Anarchy of path-disjoint free-flow games with latency functions in $\mathcal{F}$ is at most $\max\{1+\theta,{\gamma}([\mathcal{F}]_H)\}$. Such upper bound is not tight in general as that considered in Theorem \ref{thm_gen_upp_par}, but it does not require the computation of $\eta_\theta([\mathcal{F}]_H)$ for any $\theta>0$. Furthermore, if ${\gamma}([\mathcal{F}]_H)=1$ (as for weakly-monomial latency functions), we have that the Price of Anarchy is at most $1+\theta$, thus getting a simple and good upper bound for small values of $\theta$. 

\subsection{General vs Path-Disjoint/Parallel-Link Free-Flow Games}\label{gen_par_sub}
Let $\mathcal{F}$ be a class of latency functions, and let $\gamma_\infty(\mathcal{F})$ be the upper bound defined as in  Subsection~\ref{sub_int_1}. One can easily observe that $\gamma_\infty(\mathcal{F})\leq \gamma_\infty([\mathcal{F}]_H)$ in general\endnote{Let $f$ be an arbitrary latency function; we have that $f$ can be written as $f(x)=g(x)+\beta$, where $g\in [\mathcal{F}]_H$ and $\beta\geq 0$. One can easily show that $\frac{kf(k)}{f(k)(k-l)+lf(l)}=\frac{k(g(k)+\beta)}{(g(k)+\beta)(k-l)+l(g(l)+\beta)}=\frac{kg(k)+k\beta}{g(k)(k-l)+lg(l)+k\beta}\leq \frac{kg(k)}{g(k)(k-l)+lg(l)}$. Thus, by the arbitrariness of $f$, we have that $\gamma_\infty(\mathcal{F})\leq \gamma_\infty([\mathcal{F}]_H)$.}, and $\gamma_\infty(\mathcal{F})=\gamma_\infty([\mathcal{F}]_H)$ if $[\mathcal{F}]_H\subseteq \mathcal{F}$. We have the following remark (whose proof is deferred to the appendix).
\begin{remark}\label{fact_gen_dis}
$\lim_{\theta\rightarrow \infty}\max\{{\gamma}([\mathcal{F}]_H),\eta_\theta([\mathcal{F}]_H)\}=\gamma_\infty([\mathcal{F}]_H)=\lim_{\theta\rightarrow \infty}\max\{{\gamma}([\mathcal{F}]_H),\gamma_\theta([\mathcal{F}]_H)\}$.
\end{remark}
By combining Theorems \ref{thm_gen_upp} and \ref{thm_gen_upp_par} with Remark \ref{fact_gen_dis}, we get that the Price of Anarchy of $\theta$-free-flow path-disjoint (resp. parallel-link) games and that of general $\theta$-free-flow congestion games converge to the same value $\gamma_\infty([\mathcal{F}]_H)$ (for $\theta$ tending to $\infty$) if $\mathcal{F}$ is weakly (resp. strongly) diverse. However, the rate of convergence can be significantly lower in path-disjoint games. In fact, if $\mathcal{F}:=\mathcal{WM}_p$, by using the bounds provided in Theorems \ref{thm_gen_pol} and \ref{thm_gen_pol_par}, we have that $$\max\{{\gamma}([\mathcal{F}]_H),\eta_\theta([\mathcal{F}]_H)\}=\eta_\theta([\mathcal{WM}_p]_H)<\gamma_\theta([\mathcal{WM}_p]_H)=\max\{{\gamma}([\mathcal{F}]_H),\gamma_\theta([\mathcal{F}]_H)\}.$$ For instance, for $p=4$ and $\theta=1$, we get $\eta_\theta(\mathcal{WM}_p)=1.3652$ and $\gamma_\theta(\mathcal{WM}_p)=1.6994$ (see Figure~\ref{Fig: main-a} for a more detailed comparison over all values $\theta\in [0,1]$). This is an important difference with the classical setting of $\theta=\infty$, where the Price of Anarchy of general congestion games with latency functions in $\mathcal{F}$ is matched by a path-disjoint (resp. parallel-link) game, if $\mathcal{F}$ is weakly (resp. strongly) diverse\footnote{We recall from Subsection \ref{sub_int_1} that we have a stronger statement for $\theta=\infty$. Indeed, as shown by  \cite{R03} and \cite{C+04}, the Price of Anarchy of general congestion games with latency functions in $\mathcal{F}$ is matched by a  path-disjoint (resp. parallel-link) game, if $\mathcal{F}$ is not homogeneous (resp. scale-closed and not homogeneous).}. Instead, in our case the Price of Anarchy of general $\theta$-free-flow can be higher even than the one for $\theta$-free-flow path-disjoint/parallel-link games.

Finally, if $\theta=0$, by Remarks \ref{rema1} and \ref{rema2} we have that $\gamma_\theta([\mathcal{F}]_H)=\eta_\theta([\mathcal{F}]_H)=1$; thus, by Theorems~\ref{thm_gen_upp} and \ref{thm_gen_upp_par}, we get that the Price of Anarchy of $\theta$-free-flow path-disjoint (resp. parallel-link) games and that of general $\theta$-free-flow congestion games are both equal to $\gamma([\mathcal{F}]_H)$, if $\mathcal{F}$ is weakly (resp. strongly) diverse.\endnote{We have a stronger statement for $\theta=0$, too. Indeed, as shown in the appendix (Theorem \ref{thm_low_gen_3}), the Price of Anarchy of $0$-free-flow general congestion games with latency functions in $\mathcal{F}$ is matched by a $0$-free-flow path-disjoint (resp. parallel-link) game, if $[\mathcal{F}]_H\subseteq \mathcal{F}$ (resp. if $\mathcal{F}$ is scale-closed and $[\mathcal{F}]_H\subseteq \mathcal{F}$).}

We conclude that, if the considered class of latency functions is weakly (resp. strongly) diverse, the Price of Anarchy of $\theta$-free-flow path-disjoint (resp. parallel-link) games and that of general $\theta$-free-flow congestion games are equal for $\theta\in \{0,\infty\}$, but might be different for $\theta\in (0,\infty)$.

\section{Experimental Evidence for \( \theta \)-Free-Flow Time in Singapore}
\label{sec:data}

We look for experimental evidence that commuters use the heuristic presented in the introduction to guide their routing decisions. Namely, we make the conjecture that commuters consider only paths with ``length'' at most a multiplicative factor $1+\theta$ away from the shortest path taking them to their destination (where ``length'' is measured as a latency, or travel time). Does this conjecture hold in practice?

To answer, we must obtain data on the routing behavior of a sampled population. Knowing the route taken by individuals in the sample, we must be able to infer what their travel time would have been in free-flow road conditions, i.e., without anyone else on the road (the \textit{data free-flow time}). Finally, we must compute the shortest free-flow travel time on any path connecting their origin to their destination (the \textit{best free-flow time}). By comparing data and best free-flow time for each individual in the sample, we arrive at a distribution of $\theta$ over our set of trips.\footnote{Modelling assumptions and a formal definition of $\theta$ are presented in Section~\ref{sec:model}.}

In this section, we make use of the Singapore National Science Experiment dataset to understand the routing behavior of its participants. First, we introduce the dataset and our data processing methods. Second, we provide the methodology for estimating the \textit{data} free-flow trip duration of the subjects' chosen morning route, computed from the collected data. Third, we compare this measure with the \textit{best} free-flow time, optimized over all commuting paths.

\subsection{The National Science Experiment}

As part of the Smart Nation programme, the \emph{National Science Experiment} (NSE) is a nationwide project in Singapore in which over 90,000 students from primary, secondary and junior college wore a sensor, called SENSg, for up to one week per student in 2015 and 2016. The SENSg sensors collect ambient temperature, relative humidity, atmospheric pressure, light intensity, sound pressure level, and 9-degree of freedom motion data. The NSE initiative led up to the mass-production of 50,000 sensor nodes. The SENSg scans the Wi-Fi hotspots which are used to localize the sensor nodes as well as to move sensor data to a back-end server. All environment and motion values are sampled every 13 seconds using the Wi-Fi based localization system. The raw collected datapoints are then post-processed to obtain semantic data, employing state-of-the-art methods described in  \cite{Monnot2016}, \cite{monnot2017routing} and \cite{gemici2019wealth}. The semantic data covers the identification of individual trips within the discrete stream of locations, inference of the activity performed at each endpoint and transportation mode classification.

The NSE 2016 dataset contains records from 49,526 students who wore the SENSg sensor. By using a random forest algorithm \citep{Wilhelm2017}, we can identify five different modes of transportation, namely: (a) stationary; (b) walking; (c) riding a train; (d) riding a bus; and (e) riding a car. With additional information from Singapore's Land Transportation Authority, the algorithm detection covers 8 rail lines, 106 train stations, 260 bus services and 4,684 bus stops. Similarly, the 164 km of expressways and the 698 km of arterial roads in Singapore feed the algorithm to distinguish whether a subject is traveling in a car.

To ensure the quality of our empirical results, we perform a strict data cleaning process over the complete dataset. A total of 34,121 clean trips are considered, with 16,563 unique students and 89 different schools. This work focuses on morning travels of students who get to their schools from their homes. Two main reasons were considered for this choice.

First, in the following analyses, the latency, or duration of the trip, is considered as the primary ``cost'' of the subjects, discounting any other monetary cost. Morning trips typically feature subjects optimizing to minimize their latency. Evening trips are more sparse since the battery of sensor is expected to be charged at night while the subject is home. By the end of the day, if it has run out due to not being charged properly, the evening trip is not recorded. We have however in the dataset 21,065 samples for which both morning and evening trips are recorded. For these pairs, the average duration of the morning trip is 29 minutes and 6 seconds, while it is 33 minutes and 33 seconds for the evening trip.

Second, the data source---students of Singapore---may not constitute a fully representative sample of Singapore's population. However, their exposure to traffic during the morning hours---which are effectively the most congested conditions---allows us to infer properties of the system as a whole. The geographical distribution of their homes broadly correspond to the population density of Singapore, and thus provides additional confidence on the representativeness. Additionally, the number of students by school type is approximately equally distributed, hence capturing the routing behavior of subjects over a large space in Singapore.

Our dataset contains highly granular information concerning the routing decisions of the subjects. With the help of the onboard sensors in the device and the mode identification algorithm, we are able to obtain for each trip an accurate representation of its segments and their endpoints. For instance, typical segments making up a trip may be ``Walk - Car - Walk'', or ``Walk - Bus - Train - Bus - Walk''. The following study focuses on car trip segments. In this dataset \citep{monnot2017routing}, looking at the population of public transport users only, Price of Anarchy was upper bounded by 1.18. Converserly, Price of Anarchy for car users only was bounded by 1.86. Putting both populations together, Price of Anarchy was bounded by 1.34.

\subsection{Estimation of Free-Flow Time for Selected Route}

We compute a graph representation from a road map of Singapore, where each vertex is located at an intersection or a bend in the road. An edge connecting two vertices indicates the presence of a segment of road going from one vertex to the other. Edges also possess additional metadata: their physical length (in meters) as well as the road type---such as expressway, local street, arterial road, and so on.

Every edge is assigned with a cost representing how much time is needed to traverse it. This latency is obtained from edge features such as the road type and the posted speed limit on the road. For each private transportation trip segment in the dataset, we associate its origin and destination with the closest vertex in the graph. We run a shortest path algorithm to estimate the free-flow travel time of the trip segment, referred to in the following as the \textit{best free-flow time}.  This best free-flow time is compared with the \textit{data free-flow time}, or the time it would take the subject to travel the same trip segment if no one was on the road. We describe how the data free-flow time is estimated in the following paragraph.

\begin{figure}[!h]
  \centering
  \includegraphics[width=0.8\textwidth]{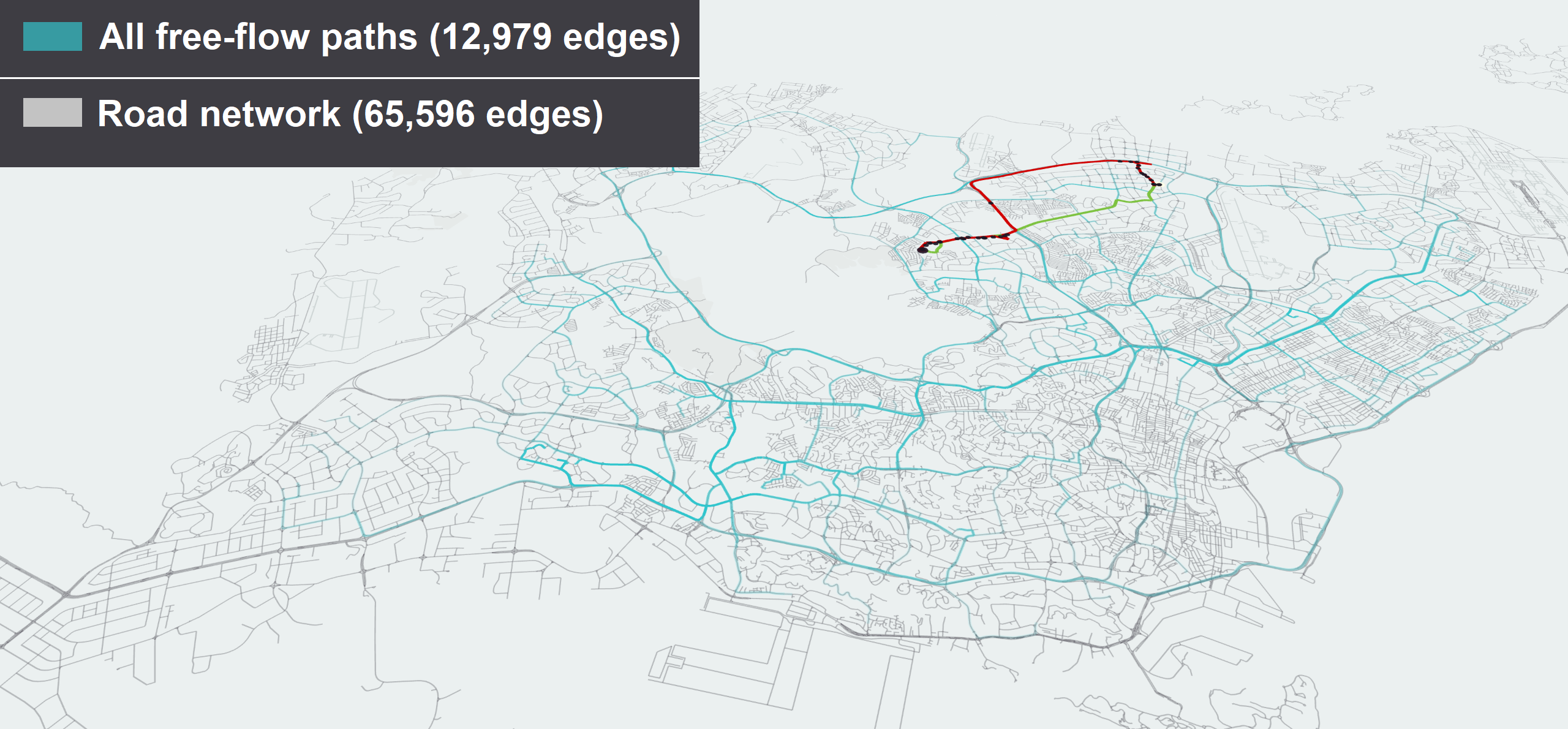}
  \caption{ For each trip segment, we find the best free-flow time and the data free-flow time. The reconstruction of the selected route uses datapoints logged along the trip. 
  }
  \label{fig:reconstruction}
\end{figure}

A segment measured by the sensor consists of a stream of geographical locations. For each datapoint, we associate the closest edge in the graph. The size of the graph (61,151 vertices and 65,596 edges) implies a lengthy lookup phase to associate the point to its closest edge. For this reason, we consider a smaller dataset of 449 car segments out of the 17,897 segments in the larger dataset.
These selected segments are well distributed across Singapore as depicted by Figure \ref{fig:reconstruction}.

The direction in which the subject traversed the edge is assigned by a heuristic based on the distance of each endpoint to the endpoints of edges preceding and following the edge under consideration. In other words, the heuristic attempts to minimize the amount of back and forth, selecting the direction that least creates deviations.

Information on the origin and destination of the trip as well as the list of directed edges traversed by the subject does not suffice. Where the sensor does not record a datapoint,\footnote{Geographical location is obtained by scanning surrounding WiFi access points. The method does not always yield accurate enough measurements, but the issue can be mitigated with proper data processing \citep{Monnot2016}.} we must provide a best guess on which edges were crossed during the trip.
\begin{itemize}
  \item For gaps of small length between two directed edges \( e_1 \) and \( e_2 \) (in that order), we compute the average speed between the two edges and drive a straight line between the target of \( e_1 \) and the source of \( e_2 \). The duration to cross this gap is obtained as the geographical distance divided by the average speed.
  \item For gaps of larger length, we run a shortest path algorithm between the target of \( e_1 \) and the source of \( e_2 \).
\end{itemize}
The data free-flow time is finally obtained as the sum of durations of redirected edges, small gaps and large gaps.

\begin{figure}[!h]
  \centering
  \includegraphics[width=0.8\textwidth]{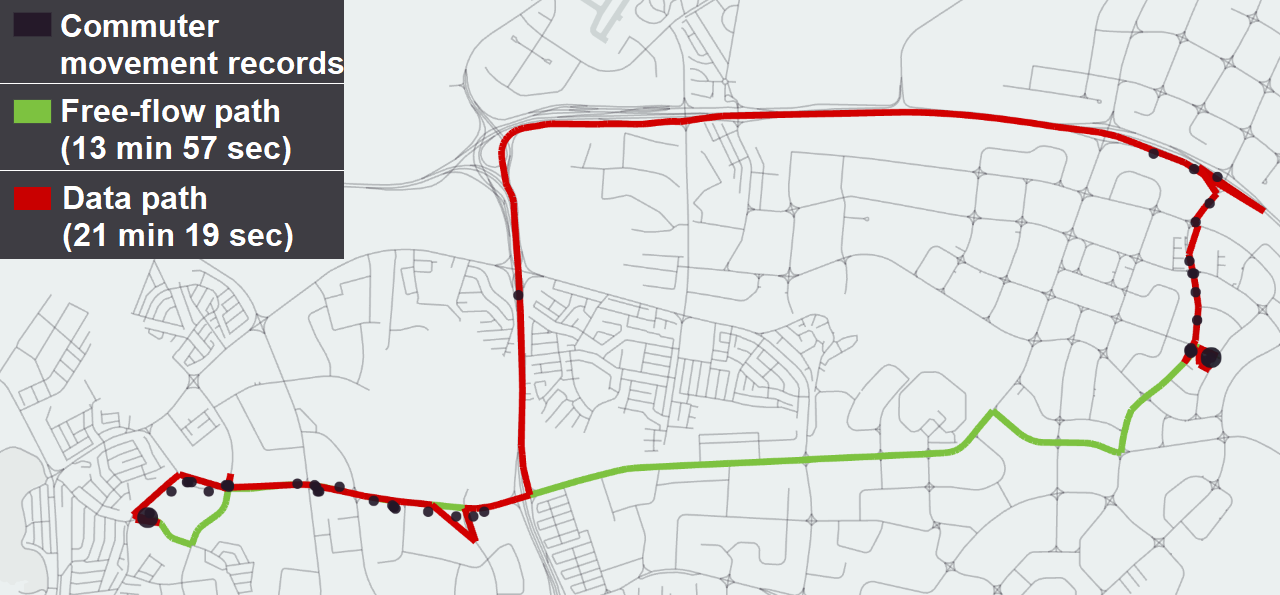}
  \caption{In green, the fastest route in free-flow condition is highlighted. The reconstructed route is in red, along which we find the data free-flow time.
  }
  \label{fig:single_agnet}
\end{figure}

\subsection{Deviation and Estimate of \( \theta \)}

\begin{figure}
  \centering
  \includegraphics[width=0.75\textwidth]{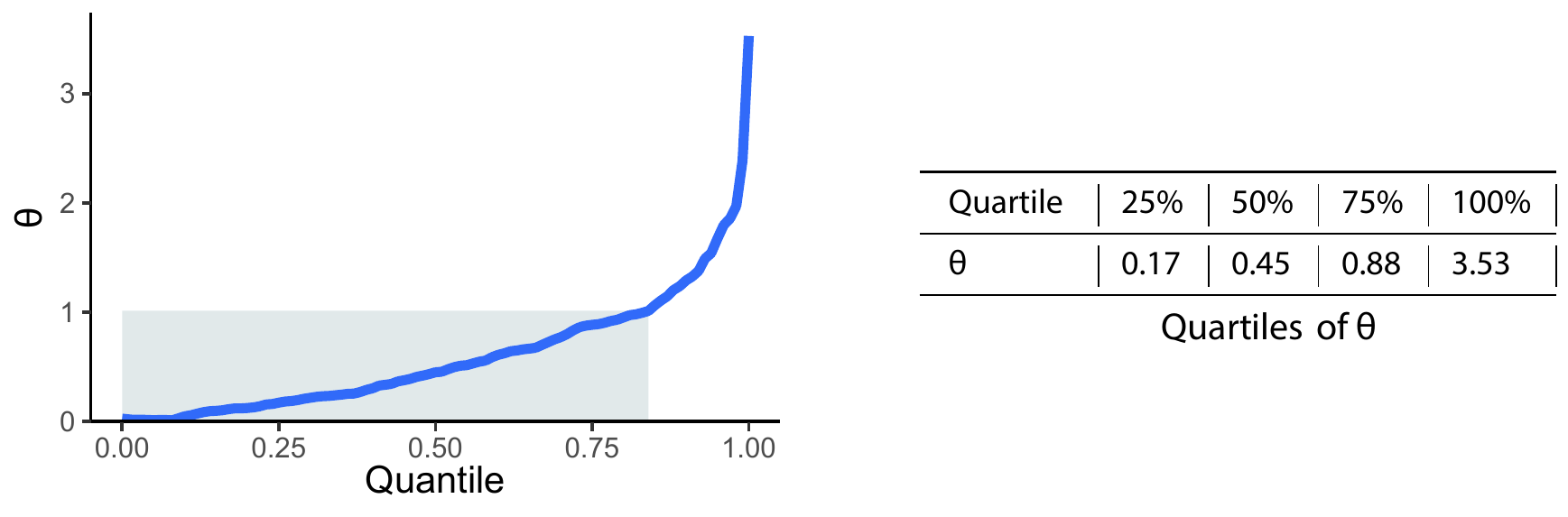}
  \caption{The deviation is measured by the ratio of the selected route free-flow time to the minimum free-flow time among all routes between the origin and the destination. Close to 80\% of the $\theta$ values are below 1, implying that the free-flow time of the selected route is rarely twice as long as the best free-flow time.}
  \label{fig:quantiles}
\end{figure}

For each trip segment, two estimates are obtained: the best free-flow time and the data free-flow time. We call \textit{deviation} the ratio between these two estimates. The deviation is strongly related to the parameter \( \theta \) we introduce in Section~\ref{sec:intro}. It measures the free-flow time difference between the best route the subject could have chosen and the route actually selected, both in a situation of no congestion. The distribution of the deviation among subjects provides a clue to estimating \( \theta \) for the routing game of Singapore. A small value of \( \theta \) yields support to the hypothesis that agents only consider routes which connect origin and destination in a straightforward manner (under no congestion) as part of their strategy set, see Figure~\ref{fig:quantiles}.

This experimental result provides justification for the upper bound of PoA estimated from the same dataset in previous work \citep{monnot2017routing}. 
This benchmark is meaningful for real road networks, as latency functions are typically estimated using affine quartic monomials \citep{bureau1964traffic}. As noted in our introduction as well as in more details in the next section, our model is based on the assumption of a uniform $\theta$ bound over the whole population. We should note that this assumption is consistent with our experimental measurements, since these measurements provide us with estimates on the lower bounds of the agents' $\theta$'s. More detailed models with a heterogeneous population/distribution of $\theta$'s is an interesting direction for future work.


\section{Concluding Remarks}
In this paper, we introduce the class of $\theta$-free-flow routing games, aiming to capture the behavior of real-world networks with a stronger assumption on edge costs than typical PoA analysis. This assumption is supported by granular data of commuters' car trips in Singapore. Indeed, the data shows evidence that agents only evaluate a small subset of their entire strategy sets to solve the routing problem. Specifically, 75\% of the agents would consider paths that are at most 88\% longer that the shortest path at free-flow. Price of anarchy analysis in  $\theta$-free-flow routing games (and variants thereof) provides much tighter Price of Anarchy guarantees that can be significantly smaller than the vanilla PoA bounds and which themselves are in better agreement with experimental investigations of Price of Anarchy  \citep{monnot2017routing}. Furthermore, we show that the Price of Anarchy in $\theta$-free-flow routing games, in general, is not independent on the network topology, differently from what happens in classical non-atomic congestion games \citep{R03}.

As a by-product of our analysis, we also determine the structure of a parallel-links game that matches the Price of Anarchy of games with homogeneous latency functions, thus solving an open problem posed by  \cite{R03}, and we tighten several bounds on the Price of Anarchy shown in \cite{BV20} for the case of affine functions, which are extended to more general latency functions, too.

We hope that this paper opens up a new direction for tighter coupling between data analytics, modelling and theory in congestion games and beyond.
Analyzing different cities as well as introducing models that take into account the difference between public and private transport seems like an exciting direction for future work.





\begin{APPENDIX}{}
\section{Omitted Proofs}
\setcounter{table}{0}
\renewcommand{\thetable}{A\arabic{table}}
\setcounter{figure}{0}
\renewcommand{\thefigure}{A\arabic{figure}}
\subsection{Tightness of the Upper Bound Shown in Theorem \ref{thm_gen_upp}}
In the following theorem, we show that the upper bound provided in Theorem \ref{thm_gen_upp} is tight for single-source network games if $\mathcal{F}$ is weakly diverse and even for load balancing games if $\mathcal{F}$ is strongly diverse.
\begin{theorem}\label{thm_low_gen_3}
Fix a value $\theta \geq 0$, a class of latency functions $\mathcal{F}$ and a value $M<\max\{\gamma([\mathcal{F}]_H),\gamma_\theta([\mathcal{F}]_H)\}$:
\begin{enumerate}[label=(\roman*)]
\item If $\gamma([\mathcal{F}]_H)\leq \gamma_\theta([\mathcal{F}]_H)$ and $\mathcal{F}$ is strongly diverse, there exists a  $\theta$-free-flow (non-symmetric) load balancing game $\LBG_\theta$ with latency functions in $\mathcal{F}$ such that $\poa(\LBG_\theta)>M$.
\item If $\gamma([\mathcal{F}]_H)\leq \gamma_\theta([\mathcal{F}]_H)$ and $\mathcal{F}$ is weakly diverse, there exists a $\theta$-free-flow single-source network congestion game $\NCG_\theta$ such that $\poa(\NCG_\theta)>M$. 
\item If $\gamma([\mathcal{F}]_H)\geq \gamma_\theta([\mathcal{F}]_H)$, $\mathcal{F}$ is scale-closed, and $[\mathcal{F}]_H\subseteq \mathcal{F}$, there exists a parallel-link game $\PLG$ (not depending on $\theta$) with latency functions in $[\mathcal{F}]_H\subseteq \mathcal{F}$ such that $\poa(\PLG)>M$.
\item If $\gamma([\mathcal{F}]_H)\geq \gamma_\theta([\mathcal{F}]_H)$ and $[\mathcal{F}]_H\subseteq \mathcal{F}$, there exists a path-disjoint network congestion game $\PNCG$ (not depending on $\theta$) with latency functions in $[\mathcal{F}]_H\subseteq \mathcal{F}$ such that $\poa(\PNCG)>M$.
\end{enumerate}
\end{theorem}
The proof of Theorem \ref{thm_low_gen_3} is based on two further theorems (Theorem \ref{thm_low_gen} and \ref{thm_low_gen_2}), whose proofs are deferred to Subsections \ref{subsub1} and \ref{subsub2}. 
\begin{theorem}\label{thm_low_gen}
Fix a value $\theta\geq 0$, a class of latency functions $\mathcal{F}$, and a value $M<\gamma_\theta([\mathcal{F}]_H)$:
\begin{enumerate}[label=(\roman*)]
\item If $\mathcal{F}$ is strongly diverse, then there exists a $\theta$-free-flow (non-symmetric) load balancing game $\LBG_\theta$ with latency functions in $\mathcal{F}$ such that $\poa(\LBG_\theta)>M$.
\item If $\mathcal{F}$ is weakly diverse only, then there exists a $\theta$-free-flow single-source network congestion game $\NCG_\theta$ such that $\poa(\NCG_\theta)>M$.
\end{enumerate}
\end{theorem}
\begin{theorem}\label{thm_low_gen_2}
Fix a class of latency functions $\mathcal{F}$ and a value $M<{\gamma}(\mathcal{F})$:
\begin{enumerate}[label=(\roman*)]
\item If $\mathcal{F}$ is scale-closed, then there exists a parallel-link game $\PLG$ with latency functions in $\mathcal{F}$ such that $\poa(\PLG)> M$.
\item If $\mathcal{F}$ is arbitrary, there exists a path-disjoint network congestion game $\PNCG$ with latency functions in $\mathcal{F}$ such that $\poa(\PNCG)>M$. 
\end{enumerate}
\end{theorem}
Given Theorems \ref{thm_low_gen} and \ref{thm_low_gen_2}, the claim of Theorem \ref{thm_low_gen_3} easily follows. Indeed, part {\em (i)} (resp. {\em (ii)}) immediately follows from part {\em (i)} (resp. {\em (ii)}) of Theorem \ref{thm_low_gen}. Furthermore, if we apply Theorem \ref{thm_low_gen_2} to the class of latency functions $[\mathcal{F}]_H$ (that is contained in $\mathcal{F}$ by hypothesis), we have that the tight bounds provided in part {\em (i)} (resp. {\em (ii)}) of Theorem  \ref{thm_low_gen_2} can be used to show part {\em (iii)} (resp. {\em (iv)}) of Theorem \ref{thm_low_gen_3}.

\subsubsection{Proof Theorem \ref{thm_low_gen}}\label{subsub1}
We first show part {\em (i)}. We do not consider the case $\theta=0$, since in such case $\gamma_\theta([\mathcal{F}]_H)=1$ (by Remark \ref{rema1}), and any congestion game has a Price of Anarchy of at least $1$. Thus, we assume that $\theta>0$. Let $k,l>0$, with $k>l$, and $f\in[\mathcal{F}]_H$ be such that $\gamma_\theta(k,l,f):=\frac{(k-l)f(k)+kf(k)\theta}{(k-l)f(k)+[(k-l)f(k)+lf(l)]\theta}>M$ (such a triple $(k,l,f)$ exists by the definition of supremum and since $M<\gamma_\theta([\mathcal{F}]_H)$). Observe that $k$ and $l$ can be chosen in such a way that $\frac{ln}{k}$ is integer for some $n\in\mathbb{N}$. Indeed, if $\frac{ln}{k}$ is not integer, we proceed as follows. First of all, observe that function $\gamma_\theta(k,t,f)$ is continuous in $t\in (0,k)$ for any fixed $k>0$ and $f\in[\mathcal{F}]_H$. Since $\gamma_\theta(k,l,f)>M$, by exploiting the continuity of $\gamma_\theta(k,t,f)$ with respect to $t$, we have that there exists a value $l'$ sufficiently close to $l$ such that $\frac{l'n}{k}$ is an integer and $\gamma_\theta(k,l',f)>M$.

To construct the lower bounding instance, we resort to a representation called {\em load balancing graph}: {\em (a)} the nodes are the resources, {\em (b)} each edge $(u,v)$ is a player having two strategies $\{u\}$ and $\{v\}$, where $u$ (resp. $v$) is called {\em the first resource} (resp. {\em the second resource}) of the considered player; {\em (c)} the weight $w_e$ of any edge $e=(u,v)\in E$ denotes the total amount of players whose feasible strategies are $\{u\}$ and $\{v\}$ only. Given an integer $m\geq 2$, let $\LBG_\theta$ be the load balancing game associated to a load balancing graph $G_\theta=(V,E)$ defined as follows: {\em (a)} the nodes of $V$ are partitioned into $m$ {\em levels}, where each level $s\in [m]$ has $n^{m-1}\left(\frac{l}{k}\right)^{m-s}$ nodes (observe that such number is an integer as $\frac{ln}{k}\in\mathbb{N}$); {\em (b)} for any level $s\in [m-1]$, there are $n^{m}\left(\frac{l}{k}\right)^{m-s}$ edges going from $s$ to $s+1$ in such a way that the out-degree of each node $u$ at level $s$ is $n$, and the in-degree of each node $v$ at level $s+1$ is $\frac{ln}{k}$; {\em (c)} $w_e=\frac{k}{n}$ for any edge $e\in E$, and the latency function of any node/resource at level $s\in [m]$ is defined as $\ell_s(x)=\alpha_s f(x)+\beta_s$, where $\alpha_s=1-(1+\theta)^{s-m}$, and $\beta_s=(1+\theta)^{s-m}f(k)$. Observe that $\LBG_\theta$ is a $\theta$-free-flow game (as $\beta_s(1+\theta)=\beta_{s+1}$ for any $s\in [m-1]$) with latency functions in $\mathcal{F}$. Lower bounding instances based on multi-level graphs have been also considered for atomic congestion games in some previous works (see \citep{bilo2017impact,BiloMV18,BiloMMV20}), but the structure of the lower bound considered here is substantially different as it is related to non-atomic games and the further $\theta$-free flow condition is taken into account. 

Leg $\sg$ and $\sg^*$ be the strategy profiles in which each player selects her first and her second resource, respectively. Observe that $k_u(\sg)=k$ (resp. $k_u(\sg^*)=l$) for any resource $u$ at level $s\in [m-1]$ (resp. $s\in [m]\setminus\{1\}$), and $k_u(\sg)=0$ for resources at level $m$ (resp. $1$). Now, we show that $\sg$ is a pure Nash equilibrium. Given $s\in [m-1]$ and a player $(u,v)$ such that $u$ is at level $s$, we get $\ell_s(k_u(\sg))=\alpha_s f(k_u(	\sg))+\beta_s=(1-(1+\theta)^{s-m})f(k)+(1+\theta)^{s-m}f(k)=f(k)=(1-(1+\theta)^{s+1-m})f(k)+(1+\theta)^{s+1-m}f(k)=\alpha_{s+1} f(k_v(\sg))+\beta_{s+1}=\ell_{s+1}(k_v(\sg))$, and this shows that $\sg$ is a pure Nash equilibrium.

We have that
\begin{align}
&{\sf SUM}(\sg)=\sum_{s=1}^{m-1}n^{m-1}\left(\frac{l}{k}\right)^{m-s}k\ell_s(k)=\sum_{s=1}^{m-1}n^{m-1}\left(\frac{l}{k}\right)^{m-s}kf(k),\label{eq_form_load}\\
&{\sf SUM}(\sg^*)=\sum_{s=2}^{m}n^{m-1}\left(\frac{l}{k}\right)^{m-s}l\ell_s(l)=\sum_{s=2}^{m}n^{m-1}\left(\frac{l}{k}\right)^{m-s}l\left[(1-(1+\theta)^{s-m})f(l)+(1+\theta)^{s-m}f(k)\right].\label{opt_form_load}
\end{align}
Given a sufficiently small $\epsilon>0$ such that $\gamma_\theta(k,l,f)>M+\epsilon$ and a sufficiently large $m$, by using (\ref{eq_form_load}) and (\ref{opt_form_load}) we get
\begin{eqnarray}
\poa(\LBG_\theta)
&\geq & \frac{{\sf SUM}(\sg)}{{\sf SUM}(\sg^*)}\nonumber\\
&=&\frac{\sum_{s=1}^{m-1}n^{m-1}\left(\frac{l}{k}\right)^{m-s}kf(k)}{\sum_{s=2}^{m}n^{m-1}\left(\frac{l}{k}\right)^{m-s}l\left[(1-(1+\theta)^{s-m})f(l)+(1+\theta)^{s-m}f(k)\right]}\nonumber\\
&=&\frac{\sum_{s=0}^{m-2}\left(\frac{k}{l}\right)^{s}kf(k)}{\left(\frac{k}{l}\right)\sum_{s=0}^{m-2}\left(\frac{k}{l}\right)^{s}l\left[(1-(1+\theta)^{s+2-m})f(l)+(1+\theta)^{s+2-m}f(k)\right]}\nonumber\\
&=&\frac{\sum_{s=0}^{m-2}\left(\frac{k}{l}\right)^{s}f(k)}{\sum_{s=0}^{m-2}\left(\frac{k}{l}\right)^{s}\left[(1-(1+\theta)^{s+2-m})f(l)+(1+\theta)^{s+2-m}f(k)\right]}\nonumber\\
&=&\frac{\left(\frac{\left(\frac{k}{l}\right)^{m-1}-1}{\frac{k}{l}-1}\right)f(k)}{\sum_{s=0}^{m-2}\left(\frac{k}{l}\right)^{s}f(l)+\sum_{s=0}^{m-2}\left(\frac{k(1+\theta)}{l}\right)^{s}(1+\theta)^{-m+2}(f(k)-f(l))}\nonumber\\
&=&\frac{\left(\frac{\left(\frac{k}{l}\right)^{m-1}-1}{\frac{k}{l}-1}\right)f(k)}{\left(\frac{\left(\frac{k}{l}\right)^{m-1}-1}{\frac{k}{l}-1}\right)f(l)+\left(\frac{\left(\frac{k(1+\theta)}{l}\right)^{m-1}-1}{\frac{k(1+\theta)}{l}-1}\right)(1+\theta)^{-m+2}(f(k)-f(l))}\nonumber\\
&> &\lim_{m\rightarrow \infty}\frac{\left(\frac{\left(\frac{k}{l}\right)^{m-1}-1}{\frac{k}{l}-1}\right)f(k)}{\left(\frac{\left(\frac{k}{l}\right)^{m-1}-1}{\frac{k}{l}-1}\right)f(l)+\left(\frac{\left(\frac{k(1+\theta)}{l}\right)^{m-1}-1}{\frac{k(1+\theta)}{l}-1}\right)(1+\theta)^{-m+2}(f(k)-f(l))}-\epsilon\nonumber\\
&=&\frac{\left(\frac{k(1+\theta)}{l}\right)^{m-1}\left(\frac{1}{\frac{k}{l}-1}\right)f(k)}{\left(\frac{k(1+\theta)}{l}\right)^{m-1}\left[\left(\frac{1}{\frac{k}{l}-1}\right)f(l)+\left(\frac{1}{\frac{k}{l}-\frac{1}{1+\theta}}\right)(1+\theta)(f(k)-f(l))\right]}-\epsilon\nonumber\\
&=&\frac{\left(\frac{1}{\frac{k}{l}-1}\right)f(k)}{\left(\frac{1}{\frac{k}{l}-1}\right)f(l)+\left(\frac{1}{\frac{k}{l}-\frac{1}{1+\theta}}\right)(f(k)-f(l))}-\epsilon\nonumber\\
&=&\frac{\left(\frac{k}{l}-\frac{1}{1+\theta}\right)f(k)}{\left(\frac{k}{l}-\frac{1}{1+\theta}\right)f(l)+\left(\frac{k}{l}-1\right)(f(k)-f(l))}-\epsilon\nonumber\\
&=&\frac{\left(k(1+\theta)-l\right)f(k)}{\left(k(1+\theta)-l\right)f(l)+\left(k-l\right)(1+\theta)(f(k)-f(l))}-\epsilon\nonumber\\
&=&\gamma_\theta(k,l,f)-\epsilon\label{load2.0}\\
&>& M+\epsilon-\epsilon\nonumber\\
&=& M,\label{load2.1}
\end{eqnarray}
and this shows part {\em (i)} of the claim.

Regarding part {\em (ii)}, we resort to similar arguments as in \citep{R03}: we reconsider the lower bounding instance of part {\em (i)}, and, by replacing each resource $e$ with a path $P$ simulating the latency function of $e$, we transform the load balancing instance in a lower bounding instance having the structure of a single-source network congestion game.
Let us consider a load balancing game $\LBG_\theta$ defined as in part {\em (i)}, such that $\poa(\LBG_\theta)>M$. Let $\theta,k,l,f,n,m$ be the parameters characterizing $\LBG_\theta$, and let $g\in\mathcal{F}$ be a latency function such that $g(x)=\beta$ for any $x>1$, for some $\beta>0$ (such a function exists as $\mathcal{F}$ is weakly diverse).

We can assume without loss of generality that $\theta$ and $f(k)/\beta$ are rational numbers. Indeed, if it is not the case, as $\gamma_\theta(k,l,f)$ is continuous with respect to variables $\theta, k, l$, and $f(x)$ is continuous in $x$, there exist $\theta'<\theta$, $k'>0$, $l'>0$, and an integer $n'\geq 1$, such that $\gamma_\theta(k',l',f)>M$, $k'>l'>0$, $\frac{l'n'}{k'}$ is integer, and $f(k')/\beta$ is rational. Thus, a load balancing instance $\LBG'_\theta$ based on values $\theta',k',l',f,n',m$ is a $\theta$-free-flow game (as $\theta'<\theta$), and verifies $\poa(\LBG'_\theta)>M$, thus we can consider $\LBG'_\theta$ in place of $\LBG_\theta$.

Now, let $\NCG_\theta$ be a single-source network congestion game constructed from $\LBG_\theta$ as follows. 
\begin{itemize}
\item Consider an undirected graph $G=(U,F)$ initially made of a unique source-node $p^*$. 
\item For any resource $e$ at level $s\in [m]$ (according to the above load-balancing graph representation) we add two consecutive directed paths $P_{e}:=(p_{e,0},p_{e,1},\ldots, p_{e,a_s})$ and $Q_e=(q_{e,0},q_{e,1},\ldots, q_{e,b_s})$ (written as sequences of consecutive nodes) with $p_{e,0}=p^*$ and $p_{e,a_s}=q_{e,0}$, and such that there exists a sufficiently large integer $h$ (not depending on the level $s$) with $a_s=\alpha_s h$ and $b_s=\beta_s h/\beta$ ($a_s$ and $b_s$ denote the number of edges of paths $P_e$ and $Q_e$, respectively), where $\alpha_s$ and $\beta_s$ are defined as in part {\em (i)}; we observe that the above integer $h$ exists, since $\theta$ and $f(k)/\beta$ are rational, and then the quantities $\alpha_s$ and $\beta_s/\beta$ are rational, too. 
\item We denote as $(u,v)$ the type of players selecting in game $\LBG_\theta$ the resources $u$ and $v$ as first and second strategy, respectively; for any type $(u,v)$, we add in $G$ a type-specific sink-node $q^*_{u,v}$, an edge from the last node of path $Q_u$ to $q^*_{u,v}$, and an edge from the last node of path $Q_v$ to $q^*_{u,v}$. 
\item For any amount of players of type $(u,v)$, we include the same amount of players in $\NCG_\theta$, and their strategies are all the simple paths from the source-node $p^*$ to the type-specific sink-node $q_{u,v}^*$; the type of such players in $\NCG_\theta$  is denoted as $(p^*,q^*_{u,v})$. For any type $(u,v)$ and $e\in \{u,v\}$, let $P_e(u,v)$ denote the simple path obtained by concatenating the paths $P_e,Q_e$, and the type-specific sink-node $q_{u,v}^*$; we observe that the unique simple paths connecting $p^*$ to $q_{u,v}^*$ are $P_u(u,v)$ and $P_{v}(u,v)$.
\item For any type $(u,v)$ and any resource $e\in \{u,v\}$, the latency function of each edge of path $P_e(u,v)$ is $f$ (resp. $g$) if the considered edge belongs to path $P_e$ (resp. if the considered edge belongs to path $Q_e$ or it coincides with the last edge of path $P_e(u,v)$).
\end{itemize}

By exploiting the construction of $\NCG_\theta$, we have that all strategy profiles $\sg'$ of game $\LBG_\theta$ can be mapped biunivocally into strategy profiles $\phi(\sg')$ via a map $\phi$ defined as follows: for each amount of players of type $(u,v)$ selecting resource $e\in \{u,v\}$ in $\sg'$, the same amount of players of type $(p^*,q^*_{u,v})$ selects path $P_{e}(u,v)$ in $\phi(\sg')$. We observe that map $\phi$ satisfies the following properties:
\begin{description}
\item[\bf Costs isomorphism:] the cost of each player of type $(p^*,q^*_{u,v})$ selecting a certain path $P_e(u,v)$ (with $e\in \{u,v\}$) in $\phi(\sg')$, is obtained by multiplying for $h$ the cost of any player of type $(u,v)$ selecting resource $e$ in $\sg'$, and then by adding the constant term $\beta$ (corresponding to the latency of the last edge of path $P_e(u,v)$);
\item[\bf Constant lenghts isomorphism:] the cost of each player of type $(p^*,q^*_{u,v})$ selecting a certain path $P_e(u,v)$ in $\phi(\sg')$, when evaluated in absence of congestion, is obtained by multiplying for $h$ the cost of any player of type $(u,v)$ selecting resource  $e$ in $\sg'$, again evaluated in absence of congestion, and then by adding the constant term $\beta$. 
\end{description}

By the constant lenghts isomorphism, we have that $\NCG_\theta$ is a $\theta$-free flow congestion games. Furthermore, by the  costs isomorphism, we have that the strategy profile $\phi(\sg)$ is a pure Nash equilibrium. Finally, for a sufficiently small $\delta$, and by fixing a sufficiently high integer $h$ in $\NCG_\theta$, we get 
\begin{align}
\poa(\NCG_\theta)&\geq\frac{{\sf SUM}(\phi(\sg))}{{\sf SUM}(\phi(\sg^*))}\label{disj1}\\
&=\frac{{\sf SUM}(\sg)h+W\beta}{{\sf SUM}(\sg^*)h+W\beta}\label{disj2}\\
&>\frac{{\sf SUM}(\sg)}{{\sf SUM}(\sg^*)}-\delta\label{disj3}\\
&>\gamma_\theta(k,l,f)-\epsilon\label{disj4}\\
&>M,\label{disj5}
\end{align}
where $W$ denotes the total amount of players of game $\NCG_\theta$, \eqref{disj1} holds since $\phi(\sg)$ is a pure Nash equilibrium, \eqref{disj2} holds because of the costs isomorphism, \eqref{disj3} holds since $h$ is sufficiently high, \eqref{disj4} and \eqref{disj5} hold because of \eqref{load2.0} and \eqref{load2.1} and since $\delta$ is sufficiently small. Thus, the same lower bound on the Price of Anarchy shown in part {\em (i)} hold for single-source network congestion games, and this concludes the proof.

\begin{remark}
The tight bound $\gamma_\theta([\mathcal{F}_H])$ continues to hold for single-source network congestion games if we simply assume that $[\mathcal{F}]_H\subset \mathcal{F}$, i.e., if $[\mathcal{F}]_H\subseteq \mathcal{F}$ and there exists $g\in \mathcal{F}$ such that $g(0)=\beta$ for some $\beta>0$. In such case, it is sufficient replacing each edge $e$ of the previous game $\NCG_\theta$ having latency constantly equal to $\beta$, with a set ${\sf PL}_e$ of parallel links having a latency function $g$ with $g(0)=\beta$, so that each player able to traverse edge $e$ in the initial game $\NCG_\theta$ can also traverse all the links of ${\sf PL}_e$ in the modified game. We observe that, all the players who selected edge $e$ in the equilibrium or the optimal configuration of the initial game, except for a small amount of them, in the new game will distribute equally on all the links of ${\sf PL}_e$, and the latency experienced on each link will be closed to $\beta$. By using these observations, one can easily show that the Price of Anarchy of the new game is close to that of the initial one. 
\end{remark}
\subsubsection{Proof of Theorem \ref{thm_low_gen_2}}\label{subsub2}
The machinery used to show Theorem \ref{thm_low_gen_2} inverts the proof arguments of Theorem \ref{thm_gen_upp} as follows: {\em (a)} we fix $M<{\gamma}(\mathcal{F})$ and we construct a dual program similar as that used in the proof of Theorem~\ref{thm_gen_upp}, and whose optimal value is at least $M$; {\em (b)} by resorting to the Strong Duality Theorem, we reconstruct a primal program {\sf LP}, similar as that defined in the proof of Theorem \ref{thm_gen_upp}, but with two variables and three primal constraints only, and whose optimal value is at least $M$ (by strong duality); {\em (c)} we give an almost-explicit representation of the optimal value of the primal program; {\em (d)} finally, we construct a lower bounding instance whose Price of Anarchy matches the above optimal value, thus providing a lower bound that can be arbitrarily closed to ${\gamma}(\mathcal{F})$ (by the arbitrariness of $M$). 

Steps (a-c) of the above sketch is formally treated in the proof of Lemma \ref{lem_low_gen_2}, and a similar machinery has been also considered in \citep{bilo2017impact,Vinci19,BiloV20} to convert upper bounds into tight lower bounds on the performance of congestion games and their variants. A substantial difference with respect to the existing approaches is the lower bounding instance considered in step (d), that is formally defined and analysed in the proof of Theorem~\ref{thm_low_gen_2}. 

Before showing Lemma \ref{lem_low_gen_2} and Theorem \ref{thm_low_gen_2}, we first give some preliminary notations and results.

Given a class $\mathcal{F}$ of latency functions and a value $x> 1$, define
\begin{align*}
&{\gamma}_{\mathcal{F},\leq}(x):=\sup_{0< k\leq l,f\in\mathcal{F}}\left(\frac{k+x(-k+l)}{l}\right)\frac{f(k)}{f(l)}=\sup_{0<t\leq 1,l>0,f\in\mathcal{F}}\left(t+x(-t+1)\right)\frac{f(tl)}{f(l)},\nonumber\\
&{\gamma}_{\mathcal{F},>}(x):=\sup_{0<l<k,f\in\mathcal{F}}\left(\frac{k+x(-k+l)}{l}	\right)\frac{f(k)}{f(l)}=\sup_{t>1,l>0,f\in\mathcal{F}}\left(t+x(-t+1)\right)\frac{f(tl)}{f(l)}.\nonumber
\end{align*}
\begin{remark}\label{rema3.0}
We have that $\gamma(\mathcal{F})=\inf_{x> 1}\max\{{\gamma}_{\mathcal{F},\leq}(x),{\gamma}_{\mathcal{F},>}(x)\}$.
\end{remark}
\begin{remark}\label{rema3}
${\gamma}_{\mathcal{F},\leq}(x)$ and ${\gamma}_{\mathcal{F},>}(x)$, as functions from $(1,\infty)$ to $\mathbb{R}_{\geq 0}\cup\{\infty\}$, are non-decreasing and non-increasing in $x$, respectively.
\end{remark}
We have the following lemma ensuring continuity properties of functions ${\gamma}_{\mathcal{F},\leq}(x)$ and ${\gamma}_{\mathcal{F},>}(x)$. 
\begin{lemma}\label{lem_cont_2}
(i) Function ${\gamma}_{\mathcal{F},\leq}(x)$ is continuous in each $x>1$. (ii) ${\gamma}_{\mathcal{F},>}(x)$ is either continuous in each $x>1$, or ${\gamma}_{\mathcal{F},>}(x)=\infty$ for any $x>1$.
\end{lemma}
\proof{Proof:}
We first show part {\em (i)}. We first observe that ${\gamma}_{\mathcal{F},\leq}(x)<\infty$ for any $x>1$. Now, we fix $x>1$ and we show the continuity of ${\gamma}_{\mathcal{F},\leq}$ in $x$. To this aim, it is sufficient providing a constant $c>0$ and $\delta\in (0,x-1)$ such that $|{\gamma}_{\mathcal{F},\leq}(x+\xi)-{\gamma}_{\mathcal{F},\leq}(x)|\leq |\xi| c$ for any $\xi \in [-\delta,\delta]$. We choose $\delta\in (0,x-1)$ arbitrarily, and we set $c:=1$; given $\xi\in [-\delta,\delta]$, we have that
\begin{align*}
&|{\gamma}_{\mathcal{F},\leq}(x+\xi)-{\gamma}_{\mathcal{F},\leq}(x)|\\
&= \Bigg|\sup_{0<t\leq 1,l>0,f\in\mathcal{F}}\left(t+(x+\xi)(-t+1)\right)\frac{f(tl)}{f(l)}-\sup_{0<t\leq 1,l>0,f\in\mathcal{F}}\left(t+x(-t+1)\right)\frac{f(tl)}{f(l)}\Bigg|\\
&\leq \Bigg|\sup_{\substack{0<t\leq 1,\\l>0,f\in\mathcal{F}}}\left(t+x(-t+1)\right)\frac{f(tl)}{f(l)}+\sup_{\substack{0<t\leq 1,\\l>0,f\in\mathcal{F}}}|\xi|(1-t)\frac{f(tl)}{f(l)}-\sup_{\substack{0<t\leq 1,\\l>0,f\in\mathcal{F}}}\left(t+x(-t+1)\right)\frac{f(tl)}{f(l)}\Bigg|\\
&=|\xi|\sup_{0<t\leq 1,l>0,f\in\mathcal{F}}(1-t)\frac{f(tl)}{f(l)}\\
&\leq |\xi|c,
\end{align*}
and this shows the continuity of ${\gamma}_{\mathcal{F},\leq}(x)$ in $x$.

Now, we show part {\em (ii)}; to this aim, we prove that, if there exists $z>1$ such that ${\gamma}_{\mathcal{F},>}(z)<\infty$, then ${\gamma}_{\mathcal{F},>}(x)$ is continuous in each $x>1$. We first give the following preliminary fact:
\begin{fact}\label{fact_cont}
Assume that there exists $z>1$ such that ${\gamma}_{\mathcal{F},>}(z)<\infty$. Then, for any $y>1$, we have that $\sup_{l>0,f\in\mathcal{F}}\frac{f(yl)}{f(l)}<\infty$. 
\end{fact}
\proof{Proof of Fact \ref{fact_cont}:} Let $z>1$ such that ${\gamma}_{\mathcal{F},>}(z)<\infty$. Assume by contradiction that there exists $y> 1$ such that $\sup_{l>0,f\in\mathcal{F}}\frac{f(yl)}{f(l)}=\infty$. We will show that ${\gamma}_{\mathcal{F},>}(z)=\infty$, and this contradicts our assumption on $z$. Let $h\in \N$ be such that $\hat{t}:=\sqrt[h]{y}<\frac{z}{z-1}$. We observe that $\hat{t}+z(-\hat{t}+1)>0$. As $\hat{t}^h=y$, we have that 
\begin{equation}\label{fact_prop1}
\frac{f(yl)}{f(l)}=\left(\frac{f(\hat{t}l)}{f(l)}\right)\cdot \left(\frac{f(\hat{t}^2l)}{f(\hat{t}l)}\right)\cdot \cdot \cdot \left(\frac{f(\hat{t}^{h-1}l)}{f(\hat{t}^{h-2}l)}\right)\cdot \left(\frac{f(\hat{t}^{h}l)}{f(\hat{t}^{h-1}l)}\right),
\end{equation}
for any $l>0$ and $f\in\mathcal{F}$. As $\sup_{l>0,f\in\mathcal{F}}\frac{f(yl)}{f(l)}=\infty$, for any arbitrary $M>0$ there exists $l>0$ and $f\in\mathcal{F}$ such that $\frac{f(yl)}{f(l)}>\left(\frac{M}{\hat{t}+z(-\hat{t}+1)}\right)^h$; thus, by \eqref{fact_prop1}, we have that there exists $j\in [h]$ such that $\frac{f(\hat{t}^{j}l)}{f(\hat{t}^{j-1}l)}>\frac{M}{\hat{t}+z(-\hat{t}+1)}$, i.e., we get 
\begin{equation}\label{fact_prop2}
(\hat{t}+z(-\hat{t}+1))\frac{f(\hat{t}\hat{l})}{f(\hat{l})}>M,
\end{equation}
for $\hat{l}:=\hat{t}^{j-1}l$. Thus, by \eqref{fact_prop2}, we get ${\gamma}_{\mathcal{F},>}(z)\geq (\hat{t}+z(-\hat{t}+1))\frac{f(\hat{t}\hat{l})}{f(\hat{l})}>M$, and by the arbitrariness of $M$, we necessarily have that ${\gamma}_{\mathcal{F},>}(x)=\infty$, that is a contradiction.
\Halmos

Now, assume that there exists $z>1$ such that ${\gamma}_{\mathcal{F},>}(x)<\infty$. We can show that ${\gamma}_{\mathcal{F},>}(x)<\infty$ for any $x>1$. Indeed, given $x>1$, we get 
\begin{align}
{\gamma}_{\mathcal{F},>}(x)&=\sup_{t>1,l>0,f\in\mathcal{F}}\left(t+x(-t+1)\right)\frac{f(tl)}{f(l)}\nonumber\\
&=\sup_{t\in \left(1,\frac{x}{x-1}\right),l>0,f\in\mathcal{F}}\left(t+x(-t+1)\right)\frac{f(tl)}{f(l)}\label{fact_prop3}\\
&\leq \sup_{l>0,f\in\mathcal{F}}\frac{f\left(\frac{x}{x-1}l\right)}{f(l)}\nonumber\\
&<\infty,\label{fact_prop4}
\end{align}
where \eqref{fact_prop3} holds since $t+x(-t+1)\leq 0$ for any $t\geq \frac{x}{x-1}$ (and then we can avoid such values of $t$ when considering the supremum), and \eqref{fact_prop4} follows by setting $y:=\frac{x}{x-1}$ in Fact \ref{fact_cont}. 

Now, we can show the continuity of function ${\gamma}_{\mathcal{F},>}$ in each $x>1$. As in part {\em (i)}, for a fixed $x>1$, it is sufficient providing a constant $c>0$ and $\delta\in (0,x-1)$ such that $|{\gamma}_{\mathcal{F},>}(x+\xi)-{\gamma}_{\mathcal{F},>}(x)|> |\xi| c$ for any $\xi \in [-\delta,\delta]$. We choose $\delta\in (0,x-1)$ arbitrarily, and we set $c:=y\sup_{l>0,f\in\mathcal{F}}\frac{f(yl)}{f(l)}$, where $y:=\frac{x-\delta}{x-\delta-1}$; by Fact \ref{fact_cont}, we have that $c$ is finite. Given $\xi\in [-\delta,\delta]$, we have that
\begin{align*}
&|{\gamma}_{\mathcal{F},>}(x+\xi)-{\gamma}_{\mathcal{F},>}(x)|\\
&=\Bigg|\sup_{t>1,l>0,f\in\mathcal{F}}\left(t+(x+\xi)(-t+1)\right)\frac{f(tl)}{f(l)}-\sup_{t>1,l>0,f\in\mathcal{F}}\left(t+x(-t+1)\right)\frac{f(tl)}{f(l)}\Bigg|\\
&=\Bigg|\sup_{t\in (1,y),l>0,f\in\mathcal{F}}\left(t+(x+\xi)(-t+1)\right)\frac{f(tl)}{f(l)}-\sup_{t\in (1,y),l>0,f\in\mathcal{F}}\left(t+x(-t+1)\right)\frac{f(tl)}{f(l)}\Bigg|\\
&\leq \Bigg|\sup_{\substack{t\in (1,y),\\l>0,f\in\mathcal{F}}}\left(t+x(-t+1)\right)\frac{f(tl)}{f(l)}+\sup_{\substack{t\in (1,y),\\l>0,f\in\mathcal{F}}}|\xi|(t-1)\frac{f(tl)}{f(l)}-\sup_{\substack{t\in (1,y),\\l>0,f\in\mathcal{F}}}\left(t+x(-t+1)\right)\frac{f(tl)}{f(l)}\Bigg|\\
&=|\xi|\sup_{t\in (1,y),l>0,f\in\mathcal{F}}(t-1)\frac{f(tl)}{f(l)}\\
&\leq |\xi|\sup_{l>0,f\in\mathcal{F}}(y-1)\frac{f(tl)}{f(l)}\\
&=|\xi|c,
\end{align*}
and this shows the continuity of ${\gamma}_{\mathcal{F},>}(x)$ in $x$, thus concluding the proof. \Halmos

By using the previous results, we get the following lemma.
\begin{lemma}\label{thm_cont}
If $\lim_{x\rightarrow 1^+}{\gamma}_{\mathcal{F},\leq}(x)<\lim_{x\rightarrow 1^+}{\gamma}_{\mathcal{F},>}(x)<\infty$ and $\lim_{x\rightarrow \infty}{\gamma}_{\mathcal{F},\leq}(x)>\lim_{x\rightarrow \infty}{\gamma}_{\mathcal{F},>}(x)$, then there exists $\hat{x}>1$ such that ${\gamma}_{\mathcal{F},\leq}(\hat{x})={\gamma}_{\mathcal{F},>}(\hat{x})=\gamma(\mathcal{F})$. 
\end{lemma}
\proof{Proof:}
Assume that $\lim_{x\rightarrow \infty}{\gamma}_{\mathcal{F},\leq}(x)>\lim_{x\rightarrow \infty}{\gamma}_{\mathcal{F},>}(x)$ and $\lim_{x\rightarrow \infty}{\gamma}_{\mathcal{F},>}(x)<\infty$. By Lemma~\ref{lem_cont_2}, we necessarily have that both functions ${\gamma}_{\mathcal{F},\leq}$ and ${\gamma}_{\mathcal{F},>}$ are continuous in each $x>1$. Thus, as $\lim_{x\rightarrow 1^+}{\gamma}_{\mathcal{F},\leq}(x)<\lim_{x\rightarrow 1^+}{\gamma}_{\mathcal{F},>}(x)<\infty$ and $\lim_{x\rightarrow \infty}{\gamma}_{\mathcal{F},\leq}(x)>\lim_{x\rightarrow \infty}{\gamma}_{\mathcal{F},>}(x)$ (by hypothesis), by Remarks  \ref{rema3.0} and \ref{rema3}, and by the Intermediate Zero Theorem, we have that there exists $\hat{x}>1$ such that ${\gamma}_{\mathcal{F},\leq}(\hat{x})={\gamma}_{\mathcal{F},>}(\hat{x})=\inf_{x>1}\max\{{\gamma}_{\mathcal{F},\leq}(\hat{x}),{\gamma}_{\mathcal{F},>}(\hat{x})\}=\gamma(\mathcal{F})$. \Halmos

The following lemma is the main ingredient to show the desired lower bound, and the proof is based on Lemma \ref{thm_cont}. Given $k_1,k_2,l_1,l_2>0$ and $f_1,f_2\in \mathcal{F}$, let 
$${\gamma}(k_1,l_1,f_1,k_2,l_2,f_2):=\frac{(l_2-k_2)f_2(k_2)k_1f_1(k_1)+(k_1-l_1)f_1(k_1)k_2f_2(k_2)}{(l_2-k_2)f_2(k_2)l_1f_1(l_1)+(k_1-l_1)f_1(k_1)l_2f_2(l_2)}.$$
\begin{lemma}\label{lem_low_gen_2}
Fix a value $M<{\gamma}(\mathcal{F})$. Then, there exist $k_1,l_1,f_1,k_2,l_2,f_2$ with $f_1,f_2\in\mathcal{F}$, $0<l_1<k_1$, and $0<k_2\leq l_2$, such that ${\gamma}(k_1,l_1,f_1,k_2,l_2,f_2)>M$. Furthermore, the above parameters can be chosen in such a way that $f_1(k_1),f_2(k_2),\frac{k_1-l_1}{l_2-k_2}\in \mathbb{Q}$ and $k_2<l_2$.
\end{lemma}
\proof{Proof:}
We observe that, if we provide some parameters $k_1,l_1,f_1,k_2,l_2,f_2$ satisfying the first part of the claim, then we can replace the real values $k_1,l_1,k_2,l_2$ with some (sufficiently close) values $k_1',l_1',k_2',l_2'$ such that  $0<l_1'<k_1'$, $0<k_2'<l_2'$, ${\gamma}(k_1',l_1',f_1,k_2',l_2',f_2)>M$ and $f_1(k_1'),f_2(k_2'),\frac{k_1'-l_1'}{l_2'-k_2'}\in \mathbb{Q}$ (this can be done since functions ${\gamma}(k_1,l_1,f_1,k_2,l_2,f_2)$, $f_1(k_1), f_2(k_2)$ and $\frac{k_1-l_1}{l_2-k_2}$ are continuous in $k_1,l_1,k_2,l_2$). Thus, in the remainder of the proof, we only show that there exist $k_1,l_1,f_1,k_2,l_2,f_2$ with $f_1,f_2\in\mathcal{F}$, $0<l_1<k_1$, and $0<k_2\leq l_2$, such that ${\gamma}(k_1,l_1,f_1,k_2,l_2,f_2)>M$.

First of all, we assume that the hypothesis of Lemma \ref{thm_cont} are satisfied; thus, there exists $\hat{x}>1$ such that ${\gamma}_{\mathcal{F},\leq}(\hat{x})={\gamma}_{\mathcal{F},>}(\hat{x})={\gamma}(\mathcal{F})>M$. Then, by the definitions of ${\gamma}_{\mathcal{F},\leq}(\hat{x})$ and ${\gamma}_{\mathcal{F},>}(\hat{x})$, there exist two triples $(k_1,l_1,f_1)$ and $(k_2,l_2,f_2)$, with $0<l_1<k_1$, $0< k_2\leq l_2$, and $f_1,f_2\in\mathcal{F}$, such that
\begin{equation}\label{form_low}
\left(\frac{k_1+\hat{x}(-k_1+l_1)}{l_1}\right)\frac{f_1(k_1)}{f_1(l_1)}\geq \left(\frac{k_2+\hat{x}(-k_2+l_2)}{l_2}\right)\frac{f_2(k_2)}{f_2(l_2)}>M.
\end{equation}
As $-k_1+l_1<0$ and $-k_2+l_2\geq 0$, we have that there exists $x^*\geq \hat{x}$ such that $\gamma^*:=\left(\frac{k_1+x^*(-k_1+l_1)}{l_1}\right)\frac{f_1(k_1)}{f_1(l_1)}=\left(\frac{k_2+x^*(-k_2+l_2)}{l_2}\right)\frac{f_2(k_2)}{f_2(l_2)}$. Then, by (\ref{form_low}), $(x^*,\gamma^*)$  is the optimal solution of the following linear program in variables $x,\gamma$:
\begin{align}
\overline{{\sf DLP}}:\ \min \ & \gamma\nonumber\\
\text{s.t.} \ & \gamma\geq \frac{k_1f_1(k_1)+x(-k_1f_1(k_1)+l_1f_1(k_1))}{l_1f_1(l_1)},\label{dual_const_low1}\\
\ & \gamma\geq \frac{k_2f_2(k_2)+x(-k_2f_2(k_2)+l_2f_2(k_2))}{l_2f_2(l_2)},\label{dual_const_low2}\\
& x\geq 0,\nonumber
\end{align}
and constraints (\ref{dual_const_low1}) and (\ref{dual_const_low2}) are tight for $x=x^*$. By considering the dual of the linear program considered above we get the following linear program in variables $\alpha_1,\alpha_2$:
\begin{align}
\overline{{\sf LP}}:\ \max \ & \alpha_1k_1f_1(k_1)+\alpha_2k_2f_2(k_2)\nonumber\\
\text{s.t.} \ & \alpha_1(k_1-l_1)f_1(k_1)+\alpha_2(k_2-l_2)f_2(k_2)\leq 0\label{lp_const_low1}\\
\ & \alpha_1l_1f_1(l_1)+\alpha_2l_2f_2(l_2)=1\label{lp_const_low2}\\
& \alpha_1,\alpha_2\geq 0.\nonumber
\end{align}
By the Strong Duality Theorem, the optimal solution of $\overline{{\sf LP}}$ has value $\gamma^*> M$. Furthermore, since $x^*>0$, by the complementary slackness conditions, we have that the optimal solution $(\alpha_1^*,\alpha_2^*)$ of $\overline{{\sf LP}}$ is such that constraint (\ref{lp_const_low1}) is tight, that, together with constraint (\ref{lp_const_low2}), gives
\begin{align*}
\alpha_1^*&=\frac{(l_2-k_2)f_2(k_2)}{(l_2-k_2)f_2(k_2)l_1f_1(l_1)+(k_1-l_1)f_1(k_1)l_2f_2(l_2)}\geq 0\\
\alpha_2^*&=\frac{(k_1-l_1)f_1(k_1)}{(l_2-k_2)f_2(k_2)l_1f_1(l_1)+(k_1-l_1)f_1(k_1)l_2f_2(l_2)}\geq 0.
\end{align*}
We conclude that, by putting $\alpha_1^*$ and $\alpha_2^*$ in the objective function of $\overline{{\sf LP}}$,  we get the optimal value $\gamma^*$ of $\overline{{\sf LP}}$, that is
\begin{align}
M<\gamma^*=\alpha_1^*k_1f_1(k_1)+\alpha_2^*k_2f_2(k_2)=\frac{(l_2-k_2)f_2(k_2)k_1f_1(k_1)+(k_1-l_1)f_1(k_1)k_2f_2(k_2)}{(l_2-k_2)f_2(k_2)l_1f_1(l_1)+(k_1-l_1)f_1(k_1)l_2f_2(l_2)},\label{form_fin_lem}
\end{align}
and this shows the claim if the hypothesis of Lemma \ref{thm_cont} are satisfied.

Now, assume that the hypothesis of Lemma \ref{thm_cont} are not satisfied. Thus, by Remark \ref{rema3}, we necessarily have that one of the following cases holds: {\em (a)} ${\gamma}_{\mathcal{F},\leq}(x)\leq {\gamma}_{\mathcal{F},>}(x)$ for any $x>1$; {\em (b)} ${\gamma}_{\mathcal{F},\leq }(x)\geq {\gamma}_{\mathcal{F},>}(x)$ for any $x>1$. 
If case {\em (a)} holds, as $\lim_{x\rightarrow \infty}{\gamma}_{\mathcal{F},\leq }(x)=\infty$, we necessarily have that there exists a sufficiently large $\hat{x}>1$ such that $M<{\gamma}_{\mathcal{F},\leq }(\hat{x})\leq {\gamma}_{\mathcal{F},> }(\hat{x})$. Thus, there exist two triples $(k_1,l_1,f_1)$ and $(k_2,l_2,f_2)$, with $0<l_1<k_1$, $0< k_2\leq l_2$, and $f_1,f_2\in\mathcal{F}$, such that the inequalities in \eqref{form_low} hold, and by proceeding from \eqref{form_low} as in the previous case, the claim follows if case {\em (a)} holds.

If case {\em (b)} holds, we have that $M<\gamma(\mathcal{F})\leq \lim_{x\rightarrow 1^+}\max\{{\gamma}_{\mathcal{F},\leq }(x),{\gamma}_{\mathcal{F},> }(x)\}=\lim_{x\rightarrow 1^+}{\gamma}_{\mathcal{F},\leq }(x)\leq \lim_{x\rightarrow 1^+}x=1$, where the second inequality follows from Remark \ref{rema3.0}. Thus, as $M<1$, one can easily observe that, by setting $k_2:=l_2:=l_1:=1$, and $k_1:=1+\delta$ for some $\delta>0$, we have that ${\gamma}(k_1,l_1,f_1,k_2,l_2,f_2)>M$, and this shows the claim if case {\em (b)} holds. 
\Halmos

Armed with the above lemma, we are ready to show Theorem \ref{thm_low_gen_2}.
\proof{Proof of Theorem \ref{thm_low_gen_2}:}
We first show part {\em (i)}. Let $k_1,l_1,f_1,k_2,l_2,f_2$ be the parameters specified in the claim of Lemma \ref{lem_low_gen_2}, i.e., such that $f_1,f_2\in\mathcal{F}$, $k_1>l_1>0$, $0<k_2< l_2$, ${\gamma}(k_1,l_1,f_1,k_2,l_2,f_2)>M$, and $f_1(k_1),f_2(k_2),\frac{k_1-l_1}{l_2-k_2}\in \mathbb{Q}$. Furthermore, let $n\geq 1$ be an integer such that $\left(\frac{k_1-l_1}{l_2-k_2}\right)n$ is integer, too (such an integer $n$ exists as $\frac{k_1-l_1}{l_2-k_2}\in\mathbb{Q}$). 

Let $\PLG$ be a parallel-link game defined as follows: {\em (a)} the set of resources $E$ is partitioned into two subsets $E^+$ and $E^-$; {\em (b)} $E^+$ contains $n$ resources having latency function defined as $\ell^+(x):=f_2(k_2)f_1(x)$, and $E^-$ contains $\left(\frac{k_1-l_1}{l_2-k_2}\right)n$ resources having latency function defined as $\ell^-(x):=f_1(k_1)f_2(x)$; {\em (c)} the total amount of players is $W:=\left(\frac{k_1l_2-k_2l_1}{l_2-k_2}\right)n$. 

Let $\sg$ (resp. $\sg^*$) be the strategy profile in which each resource of $E^+$ is selected by $k_1$ (resp. $l_1$) players and each resource of $E^-$ is selected by $k_2$ (resp. $l_2$) players; both strategy profiles are well-defined as $k_1|E^+|+k_2|E^-|=k_1n+k_2\left(\frac{k_1-l_1}{l_2-k_2}\right)n=W=l_1n+l_2\left(\frac{k_1-l_1}{l_2-k_2}\right)n=l_1|E^+|+l_2|E^-|$. One can easily observe that all resources have the same latency in $\sg$, thus $\sg$ is a pure Nash equilibrium; furthermore, we have that
\begin{align}
&{\sf SUM}(\sg)=|E^+|k_1\ell^+(k_1)+|E^-|k_2\ell^-(k_2)=n\left(k_1f_2(k_2)f_1(k_1)\right)+\left(\frac{k_1-l_1}{l_2-k_2}\right)n\left(k_2f_1(k_1) f_2(k_2)\right)\nonumber\\
&{\sf SUM}(\sg^*)=|E^+|l_1\ell^+(l_1)+|E^-|l_2\ell^-(l_2)=n\left(l_1f_2(k_2)f_1(l_1)\right)+\left(\frac{k_1-l_1}{l_2-k_2}\right)n\left(l_2f_1(k_1) f_2(l_2)\right),\nonumber
\end{align}
thus
\begin{align*}
\poa(\PLG)\geq \frac{{\sf SUM}(\sg)}{{\sf SUM}(\sg^*)}=\frac{\left[k_1f_2(k_2)f_1(k_1)+\left(\frac{k_1-l_1}{l_2-k_2}\right)k_2f_1(k_1) f_2(k_2)\right]n}{\left[l_1f_2(k_2)f_1(l_1)+\left(\frac{k_1-l_1}{l_2-k_2}\right)l_2f_1(k_1) f_2(l_2)\right]n}={\gamma}(k_1,l_1,f_1,k_2,l_2,f_2)>M,
\end{align*}
and this shows the part {\em (i)} of the claim.

Regarding part {\em (ii)}, we resort to a similar proof as in Theorem \ref{thm_low_gen}: we reconsider the lower bounding instance of part {\em (i)}, and transform it into a lower bounding instance having the structure of a path-disjoint network congestion game.
Let us consider a parallel-link game $\PLG$ defined as in part {\em (i)} such that $\poa(\PLG)>M$, and let $k_1,l_1,f_1,k_2,l_2,f_2,n$ be the parameters characterizing $\PLG$. Furthermore, let $h$ be an integer such that $a:=f_2(k_2)h$ and $b:=f_1(k_1)h$ are integers, too (such integer exists as $f_1(k_1),f_2(k_2)\in \mathbb{Q}$). 

Let $\PNCG$ be a path-disjoint network congestion game constructed from $\PLG$ as follows. 
\begin{itemize}
\item Consider an undirected graph $G=(U,F)$ initially empty, and we add in $G$ a source-node $s^*$ and a sink-node $t^*$.
\item For any resource $e$ of group $E^+$ (resp. $E^-$), we add a path $P_{e}:=(s^*,p_{e,1},\ldots, p_{e,a-1}, t^*)$ (resp. $Q_e=(s^*,q_{e,1},\ldots, q_{e,b-1},t^*)$). The latency function of each edge of paths of type $P_e$ (resp. $Q_e$) is $f_1$ (resp. $f_2$). 
\item The amount of players is the same as in $\PLG$, and their possible strategies are all the simple paths from $s^*$ to $t^*$.
\end{itemize} 
Let $\sg$ and $\sg^*$ be the strategy profiles defined in part {\em (i)}. As in part {\em (ii)} of Theorem \ref{thm_low_gen}, we have that all strategy profiles $\sg'$ of $\PLG$ can be mapped biunivocally into strategy profiles $\phi(\sg')$ via a map $\phi$ defined as follows: the amount of players selecting a resource $e\in E^+$ (resp. $e\in E^-$) in $\sg'$, is the same amount of those selecting path $P_{e}$ (resp. $Q_e$) in $\phi(\sg')$. We observe that, for any strategy profile $\sg$, the cost of each player in $\phi(\sg')$ is $h$ times the cost that the same player has in $\sg$. Thus, $\phi(\sg)$ is a pure Nash equilibrium in game $\PNCG$, and then $\poa(\PNCG)\geq \frac{{\sf SUM}(\phi(\sg))}{{\sf SUM}(\phi(\sg^*))}=\frac{{\sf SUM}(\sg)h}{{\sf SUM}(\sg^*)h}>M$, where the last inequality comes from the last proof steps in part {\em (i)}.
\subsection{Proof of Corollary \ref{cor1}}
To show that $\gamma(\mathcal{F})$ is an upper bound on the Price of Anarchy, it suffices fixing an arbitrary congestion games $\CG$, and considering the same linear program {\sf LP} as that considered in the proof of Theorem \ref{thm_gen_upp}, but with the following differences: {\em (a)} $\ell_e(x)$ is equal to $\alpha_e\cdot f_e$ for some $f_e\in\mathcal{F}$ (we observe that the eventual constant $\beta_e\geq 0$ is absorbed by the latency $f_e$, that now is not necessarily homogeneous), and {\em (b)} we remove variables $\beta_e$s and constraint \eqref{primcost2} (as the $\theta$-free-flow condition is not required). We have that the optimal value of the above linear program is an upper bound the Price of Anarchy of $\CG$. Furthermore, we observe that dual of such linear program is equal to ${\sf DLP}(\mathcal{T}_E)$, but without constraint \eqref{dualcost2} and variable $y$, and where each function $f_e$ belongs to $\mathcal{F}$. Thus, by reconsidering the proof steps (\ref{upp_form_1.000}-\ref{upp_form_1}), one can show that $\gamma(\mathcal{F})$ is the desired upper bound.

The fact that $\gamma(\mathcal{F})$ is a tight lower bound (under the considered assumptions on $\mathcal{F}$), immediately follows from Theorem \ref{thm_low_gen_2}.
\subsection{Proof of Lemma \ref{lemma-thm-2}}
By substituting $\beta_u:=\beta_v/(1+\theta)$ and $\alpha_u:=(\alpha_vf_v(k_v)+(1-1/(1+\theta))\beta_v)/f_u(k_u)$ in the definition of $F(\alpha_u,\beta_u,\alpha_v,\beta_v)$, we get
\begin{align}
&F(\alpha_u,\beta_u,\alpha_v,\beta_v)\nonumber\\
&=\frac{\left(\frac{k_u}{k_u-l_u}+\frac{k_v}{l_v-k_v}\right)(\alpha_v f_v(k_v)+\beta_v)}{\frac{l_u}{k_u-l_u}\left(\left(\frac{\alpha_vf_v(k_v)+\left(1-\frac{1}{1+\theta}\right)\beta_v}{f_u(k_u)}\right)f_u(l_u)+\frac{\beta_v}{1+\theta}\right)+\frac{l_v}{l_v-k_v}\left(\alpha_vf_v(l_v)+\beta_v\right)}\nonumber\\
&=\frac{\alpha_v\overbrace{\left(\left(\frac{k_u}{k_u-l_u}+\frac{k_v}{l_v-k_v}\right)f_v(k_v)\right)}^{N_\alpha}+\beta_v\overbrace{\left(\frac{k_u}{k_u-l_u}+\frac{k_v}{l_v-k_v}\right)}^{N_\beta}}{\alpha_v\overbrace{\left(\frac{l_u f_v(k_v)f_u(l_u)}{(k_u-l_u)f_u(k_u)}+\frac{l_vf_v(l_v)}{l_v-k_v}\right)}^{D_\alpha}+\beta_v\overbrace{\left(\frac{l_u}{k_u-l_u}\left(\frac{\left(1-\frac{1}{1+\theta}\right)f_u(l_u)}{f_u(k_u)}+\frac{1}{1+\theta}\right)+\frac{l_v}{l_v-k_v}\right)}^{D_\beta}}\nonumber\\
&\leq \max\left\{\frac{\overbrace{\left(\frac{k_u}{k_u-l_u}+\frac{k_v}{l_v-k_v}\right)f_v(k_v)}^{N_\alpha}}{\overbrace{\frac{l_u f_v(k_v)f_u(l_u)}{(k_u-l_u)f_u(k_u)}+\frac{l_vf_v(l_v)}{l_v-k_v}}^{D_\alpha}},\frac{\overbrace{\frac{k_u}{k_u-l_u}+\frac{k_v}{l_v-k_v}}^{N_\beta}}{\overbrace{\frac{l_u}{k_u-l_u}\left(\frac{\left(1-\frac{1}{1+\theta}\right)f_u(l_u)}{f_u(k_u)}+\frac{1}{1+\theta}\right)+\frac{l_v}{l_v-k_v}}^{D_\beta}}\right\}\nonumber\\
&= \max\left\{\overbrace{\frac{(l_v-k_v)f_v(k_v)k_uf_u(k_u)+(k_u-l_u)f_u(k_u)k_vf_v(k_v)}{(l_v-k_v)f_v(k_v)l_uf_u(l_u)+(k_u-l_u)f_u(k_u)l_vf_v(l_v)}}^{N_\alpha/D_\alpha},\overbrace{\frac{\frac{k_u}{k_u-l_u}+\frac{k_v}{l_v-k_v}}{\frac{l_u}{k_u-l_u}\left(\frac{\theta f_u(l_u)+f_u(k_u)}{(1+\theta) f_u(k_u)}\right)+\frac{l_v}{l_v-k_v}}}^{N_\beta/D_\beta}\right\}.\label{par_lin_1}
\end{align}
Now, we exploit the following facts:
\begin{fact}\label{fact_0par}
We have that
$$\frac{N_\alpha}{D_\alpha}=\frac{(l_2-k_2)f_2(k_2)k_1f_1(k_1)+(k_1-l_1)f_1(k_1)k_2f_2(k_2)}{(l_2-k_2)f_2(k_2)l_1f_1(l_1)+(k_1-l_1)f_1(k_1)l_2f_2(l_2)}\leq {\gamma}([\mathcal{F}]_H)$$ for any $k_1,l_1,f_1,k_2,l_2,f_2$ such that $k_1>l_1\geq 0$, $l_2>k_2\geq 0$, and $f_1,f_2\in [\mathcal{F}]_H$.
\end{fact}
\proof{Proof:}
Reconsider the proof of Lemma \ref{lem_low_gen_2} and the related notation. As shown in \eqref{form_fin_lem}, we have that 
$$\gamma^*:=\frac{(l_2-k_2)f_2(k_2)k_1f_1(k_1)+(k_1-l_1)f_1(k_1)k_2f_2(k_2)}{(l_2-k_2)f_2(k_2)l_1f_1(l_1)+(k_1-l_1)f_1(k_1)l_2f_2(l_2)}$$ 
is the optimal value of the linear program $\overline{\sf LP}$ (defined in the proof of Lemma \ref{lem_low_gen_2}) parametrized by $k_1,l_1,f_1,k_2,l_2,f_2$. We have that the dual of $\overline{\sf LP}$ is the linear program $\overline{\sf DLP}$ (defined in the proof of Lemma \ref{lem_low_gen_2}), and the optimal value of $\overline{\sf DLP}$ is upper bounded by ${\gamma}([\mathcal{F}]_H)$ (this  fact can be shown by exploiting similar inequalities as in (\ref{upp_form_1.000}-\ref{upp_form_1})). Then, by the Weak Duality Theorem, ${\gamma}([\mathcal{F}]_H)$ is an upper bound on $\gamma^*$, and this shows the claim. \Halmos
\begin{fact}\label{fact_1par}
We have that
\begin{equation}\label{eq_fact_par}
\frac{N_\beta}{D_\beta}=\frac{\frac{k_u}{k_u-l_u}+\frac{k_v}{l_v-k_v}}{\frac{l_u}{k_u-l_u}\left(\frac{\theta f_u(l_u)+f_u(k_u)}{(1+\theta) f_u(k_u)}\right)+\frac{l_v}{l_v-k_v}}\leq \eta_\theta([\mathcal{F}]_H).
\end{equation}
\end{fact}
\proof{Proof:}
Let $a:=\frac{k_u}{k_u-l_u}$, $b:=\frac{l_u}{k_u-l_u}\left(\frac{\theta f_u(l_u)+f_u(k_u)}{(1+\theta) f_u(k_u)}\right)\leq \frac{l_u}{k_u-l_u}<a$, and $t:=\frac{k_v}{l_v}\in [0,1)$, so that the ratio $N_\beta/D_\beta$ is equal to $\frac{a+t/(1-t)}{b+1/(1-t)}$. As $a>b$, by standard arguments of calculus, one can show that $\frac{a+t/(1-t)}{b+1/(1-t)}$ is maximized by $t=0$, i.e., $\frac{a+t/(1-t)}{b+1/(1-t)}\leq \frac{a}{b+1}$. Thus,
\begin{equation}
\frac{a+\frac{t}{1-t}}{b+\frac{1}{1-t}}\leq \frac{a}{b+1}=\frac{k_uf_u(k_u)+k_uf_u(k_u)\theta}{k_uf_u(k_u)+[(k_u-l_u)f_u(k_u)+l_uf_u(l_u)]\theta}\leq \eta_\theta([\mathcal{F}]_H).
\end{equation}
\Halmos

By using the previous facts, we get that (\ref{par_lin_1}) is upper bounded by $\max\left\{{\gamma}([\mathcal{F}]_H),\eta_\theta([\mathcal{F}]_H)\right\}$, and this shows the claim.

\subsection{Upper Bound for Path-disjoint Network Games in Theorem \ref{thm_gen_upp_par}}
Let $\overline{\mathcal{F}}:=\{g:g(x)=\sum_{r=1}^s f_r(x)\ \forall x>0, f_1,\ldots, f_s\in\mathcal{F}, s\geq 1\}$, i.e., $\overline{\mathcal{F}}$ is the smallest class of latency functions containing $\mathcal{F}$ and closed under sums of latency functions. Let $\PNCG_\theta$ be a $\theta$-free-flow path-disjoint network congestion game with latency functions in $\mathcal{F}$. For any  path $P$ (selectable as players' strategy) of $\PNCG_\theta$, we replace $P$ with a unique resource $e_P$ having latency function defined as $\overline{\ell}_{e_P}(x):=\sum_{e\in P}\ell_{e}(x)$, where $\ell_e$ is the latency function of edge $e$ in game $\PNCG_\theta$. By construction of  $\overline{\mathcal{F}}$, we have that $\overline{\ell}_{e_P}\in\overline{\mathcal{F}}$ for any path $P$ of $\PNCG_\theta$. Thus, the resulting game can be seen as a $\theta$-free-flow parallel-link game with latency functions in $\overline{\mathcal{F}}$, and by Theorem \ref{thm_gen_upp_par}, its Price of Anarchy is at most $\max\{\eta_\theta([\overline{\mathcal{F}}]_H), {\gamma}([\overline{\mathcal{F}}]_H)\}$. We observe that:
\begin{align*}
\eta_\theta([\overline{\mathcal{F}}]_H)&=\sup_{k>l>0,s\geq 1, f_1,\ldots, f_s\in[\mathcal{F}]_H}\frac{k\sum_{r=1}^sf_r(k)+k\sum_{r=1}^sf_r(k)\theta}{k\sum_{r=1}^sf_r(k)+[(k-l)\sum_{r=1}^sf_r(k)+l\sum_{r=1}^sf_r(l)]\theta}\\
&=\sup_{k>l>0,s\geq 1, f_1,\ldots, f_s\in[\mathcal{F}]_H}\frac{\sum_{r=1}^s\left[kf_r(k)+kf_r(k)\theta\right]}{\sum_{r=1}^s\left[kf_r(k)+[(k-l)f_r(k)+lf_r(l)]\theta\right]}\\
&=\sup_{k>l>0,f\in[\mathcal{F}]_H}\frac{kf(k)\theta}{kf(k)\theta+(lf(l)-lf(k))(\theta-1)}\\
&=\eta_\theta([\mathcal{F}]_H),
\end{align*}
and with analogue proof arguments, we get ${\gamma}([\overline{\mathcal{F}}]_H)={\gamma}([\mathcal{F}]_H)$. We conclude that $\poa(\PNCG_\theta)\leq \max\{\eta_\theta([\overline{\mathcal{F}}]_H), {\gamma}([\overline{\mathcal{F}}]_H)\}=\max\{\eta_\theta([\mathcal{F}]_H), {\gamma}([\mathcal{F}]_H)\}$, thus showing the claim.

\subsection{Tightness of the Upper Bound Shown in Theorem \ref{thm_gen_upp_par}}
\begin{theorem}\label{thm_low_par}
Fix a value $\theta\geq 0$, a class of latency functions $\mathcal{F}$ and a value $M<\max\{\gamma([\mathcal{F}]_H),\eta_\theta([\mathcal{F}]_H)\}$. 
\begin{enumerate}[label=(\roman*)]
\item If $\gamma([\mathcal{F}]_H)\leq \eta_\theta([\mathcal{F}]_H)$ and $\mathcal{F}$ is strongly diverse, there exists a  $\theta$-free-flow parallel-link game $\PLG_\theta$ with latency functions in $\mathcal{F}$ such that $\poa(\PLG_\theta)>M$.
\item If $\gamma([\mathcal{F}]_H)\leq \eta_\theta([\mathcal{F}]_H)$ and $\mathcal{F}$ is weakly diverse, there exists a $\theta$-free-flow path-disjoint network congestion game $\PNCG_\theta$ with latency functions in $\mathcal{F}$ such that $\poa(\PNCG_\theta)>M$.
\item If $\gamma([\mathcal{F}]_H)\geq \eta_\theta([\mathcal{F}]_H)$, $\mathcal{F}$ is scale-closed, and $[\mathcal{F}]_H\subseteq \mathcal{F}$, there exists a parallel-link game $\PLG$ (not depending on $\theta$) with latency functions in $[\mathcal{F}]_H\subseteq \mathcal{F}$ such that $\poa(\PLG)>M$.
\item If $\gamma([\mathcal{F}]_H)\geq \eta_\theta([\mathcal{F}]_H)$ and $[\mathcal{F}]_H\subseteq \mathcal{F}$, there exists a path-disjoint network congestion game $\PNCG$ (not depending on $\theta$) with latency functions in $[\mathcal{F}]_H\subseteq \mathcal{F}$ such that $\poa(\PNCG)>M$.
\end{enumerate}
\end{theorem}
\proof{Proof:}
We only show claims {\em (i)} and {\em (ii)}, as Theorem \ref{thm_low_gen_3} already guarantees the existence of parallel-link and path-disjoint games satisfying claims {\em (iii)} and {\em (iv)}. 

We first show part {\em (i)}. Fix $M<\eta_\theta([\mathcal{F}]_H)$, and let $k,l,f$ such that $f\in [\mathcal{F}]_H$, $k>l>0$, and $\eta_\theta(k,l,f):=\frac{kf(k)+kf(k)\theta}{kf(k)+[(k-l)f(k)+lf(l)]\theta}>M$. Consider a parallel-link game $\PLG_\theta$ with two resources $u$ and $v$ and an amount of players equal to $k$, where $u$ and $v$ have latency function defined as $\ell_u(x)=\theta f(x)+f(k)$ and $\ell_v(x)=(1+\theta) f(k)$, respectively. Observe that $\ell_u(0)(1+\theta)=\ell_v(0)$, thus $\PLG_\theta$ is a $\theta$-free-flow game.

Let $\sg$ (resp. $\sg^*$) be the strategy profile such that $k$ (resp. $l$) players select resource $u$, and $0$ (resp. $k-l$) players select resource $v$. One can easily observe that $\ell_u(k_u(\sg))=(1+\theta) f(k)=\ell_v(k_v(\sg))$, thus $\sg$ is a pure Nash equilibrium. We have that
\begin{align}
{\sf SUM}(\sg)&=k\ell_u(k)=k f(k)+\theta kf(k)\nonumber
\end{align}
and
\begin{align}
{\sf SUM}(\sg^*)&=l\ell_u(l)+(k-l)\ell_v(k-l)\nonumber\\
&=lf(l)\theta+lf(k)+(k-l)f(k)(1+\theta)\nonumber\\
&=kf(k)+[(k-l)f(k)+lf(l)]\theta,\nonumber
\end{align}
thus 
\begin{equation}\label{last_par}
\poa(\PLG_\theta)\geq \frac{{\sf SUM}(\sg)}{{\sf SUM}(\sg^*)}=\frac{kf(k)(1+\theta)}{kf(k)+[(k-l)f(k)+lf(l)]\theta}=\eta_\theta(k,l,f)>M,
\end{equation}
and this shows the claim of part~{\em (i)}.

To show part {\em (ii)}, it is sufficient considering similar arguments as in part {\em (ii)} of Theorems \ref{thm_low_gen} and \ref{thm_low_gen_2}.
Let $k$, $l$ and $f$ be defined as in part {\em (i)}, and let $g\in\mathcal{F}$ be a constant latency function defined as $g(x):=\beta>0$. Analogously to Theorems \ref{thm_low_gen} and \ref{thm_low_gen_2}, one can assume without loss of generality that $f(k)/\beta$ and $\theta$ are rational numbers.

Let $\PNCG_\theta$ be a path-disjoint network congestion game constructed from $\PLG_\theta$ as follows. 
\begin{itemize}
\item Consider an undirected graph $G=(U,F)$ initially empty, and we add in $G$ a source-node $s^*$ and a sink-node $t^*$.
\item Let $h\in\mathbb{N}$ be such that $a:=\theta h$, $b:=f(k)h/\beta$, and $c:=(1+\theta) f(k)h/\beta$ are integers (such integer $h$ exists as $f(k)/\beta$ and $\theta$ are rational numbers). 
\item We add in $G$ a path $P_u$ made of two concatenated sub-paths $Q_u=(q_{u,0}:=s^*,q_{u,1},\ldots, q_{u,a})$ and $R_u=(r_{u,0}:=q_{u,a},r_{u,1},\ldots, r_{u,b}:=t^*)$; we also add a path $P_v:=(p_{v,0}:=s^*,p_{v,1},\ldots, p_{v,c}:=t^*)$. 
\item The latency of each edge in path $Q_u$ (resp. $R_u$ and $P_v$) is $f$ (resp. $g$).
\item The total amount of players is the same as in $\PLG_\theta$, and their possible strategies are all the simple paths from $s^*$ to $t^*$ (i.e., paths $P_u$ and $P_v$).
\end{itemize} 
Let $\sg$ and $\sg^*$ be the strategy profiles defined in part {\em (i)}. Analogously to part {\em (ii)} of Theorems \ref{thm_low_gen} and \ref{thm_low_gen_2}, we have that all the strategy profiles $\sg'$ of $\PLG_\theta$ can be mapped biunivocally into strategy profiles $\phi(\sg')$ of $\PNCG_\theta$ via a map $\phi$ defined as follows: the amount of players selecting resource $e\in \{u,v\}$ in $\sg'$ is the same amount of those selecting path $P_{e}$ in $\phi(\sg')$. We observe that, for any strategy profile $\sg'$, the cost of each player in $\phi(\sg')$ is $h$ times the cost that the same player has in $\sg'$; the same fact holds relatively to the players' costs evaluated in absence of congestion. Thus, $\phi(\sg)$ is a pure Nash equilibrium in game $\PNCG_\theta$, $\PNCG_\theta$ is a $\theta$-free-flow game, and then $\poa(\PNCG)\geq \frac{{\sf SUM}(\phi(\sg))}{{\sf SUM}(\phi(\sg^*))}=\frac{{\sf SUM}(\sg)h}{{\sf SUM}(\sg^*)h}>M$, where the last inequality comes from \eqref{last_par}. 
\Halmos

\subsection{Proof of Theorem \ref{thm_gen_pol}}
To show the claim, by Theorem \ref{thm_gen_upp}, and since $\mathcal{P}_{p,q}$ is strongly diverse, it suffices computing the values of $\gamma_\theta([\mathcal{P}_{p,q}]_H)$ and ${\gamma}([\mathcal{P}_{p,q}]_H)$. We get
\begin{eqnarray}
\gamma_\theta([\mathcal{P}_{p,q}]_H)
&=&\sup_{k>l>0,f\in[\mathcal{P}_{p,q}]_H}\frac{(k-l)f(k)+kf(k)\theta}{(k-l)f(k)+[(k-l)f(k)+lf(l)]\theta}\nonumber\\
&=&\sup_{k>l>0,(\alpha_q,\alpha_{q+1},\ldots,\alpha_p)>0}\frac{\left(\sum_{d=q}^p\alpha_d k^d\right)(k(1+\theta)-l)}{\left(\sum_{d=q}^p\alpha_d k^d\right)(k-l)(1+\theta)+l\left(\sum_{d=q}^p\alpha_d l^d\right)\theta}\nonumber\\
&=&\sup_{k>l>0,(\alpha_q,\alpha_{q+1},\ldots,\alpha_p)>0}\frac{\sum_{d\in [p]}\alpha_d\left(k^d(k(1+\theta)-l)\right)}{\sum_{d=q}^p\alpha_d\left(k^d(k-l)(1+\theta)+l^{d+1}\theta\right)}\nonumber\\
&=&\max_{d\in [p]\setminus [q-1]}\sup_{k>l>0}\frac{k^d(k(1+\theta)-l)}{k^d(k-l)(1+\theta)+l^{d+1}\theta}\nonumber\\
&=&\max_{d\in [p]\setminus [q-1]}\sup_{t>1}\frac{t^d(t(1+\theta)-1)}{t^d(t-1)(1+\theta)+\theta}\nonumber\\
&=&\max_{d\in [p]\setminus [q-1]}\sup_{t>1}\frac{t^p(t(1+\theta)-1)}{t^p(t-1)(1+\theta)+t^{p-d}\theta}\nonumber\\
&=&\sup_{t>1}\frac{t^p(t(1+\theta)-1)}{t^p(t-1)(1+\theta)+\theta}\nonumber\\
&=&\sup_{t>1}\frac{t^{p+1}(1+\theta)-t^p}{t^{p+1}(1+\theta)-t^p(1+\theta)+\theta},\nonumber
\end{eqnarray}
thus obtaining (\ref{pol1}).

Now, we compute ${\gamma}([\mathcal{P}_{p,q}]_H)$. We recall the functions $\gamma_{\leq,\mathcal{F}}$ and $\gamma_{>,\mathcal{F}}$ defined in Subsection \ref{subsub2}, and we first show that there exist $x_0$ and $x_1$ with $1<x_0\leq x_1$ such that $\gamma_{\leq,[\mathcal{P}_{p,q}]_H}(x_0)\leq \gamma_{>,[\mathcal{P}_{p,q}]_H}(x_0)$ and $\gamma_{\leq,[\mathcal{P}_{p,q}]_H}(x_1)\geq \gamma_{>,[\mathcal{P}_{p,q}]_H}(x_1)$; similarly to Lemma \ref{thm_cont}, this fact will imply the existence of a value $\hat{x}\in [x_0,x_1]$ such that $\gamma_{[\mathcal{P}_{p,q}]_H,\leq}(\hat{x})=\gamma_{[\mathcal{P}_{p,q}]_H,>}(\hat{x})=\gamma([\mathcal{P}_{p,q}]_H)$.

Let $x_0:=q+1$ and $x_1:=p+1$. Given $x\in [x_0,x_1]$, we have that:
\begin{eqnarray}
{\gamma}_{[\mathcal{P}_{p,q}]_H,\leq}(x)
&=&\sup_{0<t\leq 1,l>0,f\in[\mathcal{P}_{p,q}]_H}\left(t+x(-t+1)\right)\frac{\sum_{d=q}^p\alpha_d (tl)^d}{\sum_{d=q}^p\alpha_d l^d}\nonumber\\
&=&\sup_{0<t\leq 1}\left(t+x(-t+1)\right)t^q\nonumber\\
&=&\frac{q^q x^{q+1}}{(q+1)^{q+1}(x-1)^{q}},\label{pol_form_2}
\end{eqnarray}
and
\begin{eqnarray}
\gamma_{[\mathcal{P}_{p,q}]_H,>}(x)
&=&\sup_{t>1,l>0,f\in[\mathcal{P}_{p,q}]_H}\left(t+x(-t+1)\right)\frac{\sum_{d=q}^p\alpha_d (tl)^d}{\sum_{d=q}^p\alpha_d l^d}\nonumber\\
&=&\sup_{t>1}\left(t+x(-t+1)\right)t^p\nonumber\\
&=&\frac{p^p x^{p+1}}{(p+1)^{p+1}(x-1)^{p}},\label{pol_form_1}
\end{eqnarray}
where \eqref{pol_form_2} and \eqref{pol_form_1} follow from standard calculations. We observe that ${\gamma}_{[\mathcal{P}_{p,q}]_H,\leq} (x_0)=1\leq{\gamma}_{[\mathcal{P}_{p,q}]_H,>} (x_0)$ and ${\gamma}_{[\mathcal{P}_{p,q}]_H,\leq} (x_1)\geq 1={\gamma}_{[\mathcal{P}_{p,q}]_H,>} (x_1)$. Since both functions ${\gamma}_{[\mathcal{P}_{p,q}]_H,\leq}$ and ${\gamma}_{[\mathcal{P}_{p,q}]_H,>}$ are continuous in each $x>1$ (by Lemma~\ref{lem_cont_2}), by Remarks \ref{rema3.0} and \ref{rema3}, and by the Intermediate Zero Theorem, we necessarily get that there exists $\hat{x}\in [x_0,x_1]$ with $\gamma_{[\mathcal{P}_{p,q}]_H,\leq}(\hat{x})=\gamma_{[\mathcal{P}_{p,q}]_H,>}(\hat{x})=\gamma([\mathcal{P}_{p,q}]_H)$.

 If $q=p$, we necessarily have that $\hat{x}=p+1$, and then $\gamma([\mathcal{P}_{p,q}]_H)=\gamma_{[\mathcal{P}_{p,q}]_H,\leq}(\hat{x})=\gamma_{[\mathcal{P}_{p,q}]_H,>}(\hat{x})=1$, thus (\ref{pol2}) holds. If $p>q$, by using the characterizations of $\gamma_{[\mathcal{P}_{p,q}]_H,\leq}(\hat{x})$ and $\gamma_{[\mathcal{P}_{p,q}]_H,>}(\hat{x})$ exploited in (\ref{pol_form_2}) and (\ref{pol_form_1}), and by using equality $\gamma_{[\mathcal{P}_{p,q}]_H,\leq}(\hat{x})=\gamma_{[\mathcal{P}_{p,q}]_H,>}(\hat{x})$, we have that $\hat{x}$ necessarily verifies $
\frac{q^q \hat{x}^{q+1}}{(q+1)^{q+1}(\hat{x}-1)^{q}}=\frac{p^p \hat{x}^{p+1}}{(p+1)^{p+1}(\hat{x}-1)^{p}}$, that is $
\hat{x}=\sqrt[p-q]{\frac{(p+1)^{p+1}q^q}{(q+1)^{q+1}p^p}}/\left(\sqrt[p-q]{\frac{(p+1)^{p+1}q^q}{(q+1)^{q+1}p^p}}-1\right)$; finally, by substituting the obtained value $\hat{x}$ in \eqref{pol_form_2} (or \eqref{pol_form_1}), we get (\ref{pol2}).

\subsection{Proof of Theorem \ref{thm_gen_pol_par}}
To show the claim, by Theorem \ref{thm_gen_upp_par}, and since $\mathcal{P}_{p,q}$ is strongly diverse, it suffices computing the values of $\eta_\theta([\mathcal{P}_{p,q}]_H)$ and ${\gamma}([\mathcal{P}_{p,q}]_H)$. ${\gamma}([\mathcal{P}_{p,q}]_H)$ has been computed in Theorem \ref{thm_gen_pol}, thus it is sufficient computing the value of $\eta_\theta([\mathcal{F}]_H)$. We have that:
\begin{eqnarray}
\eta_\theta([\mathcal{P}_{p,q}]_H)
&=&\sup_{k>l>0,f\in\mathcal{G}}\frac{kf(k)+kf(k)\theta}{kf(k)+[(k-l)f(k)+lf(l)]\theta}\nonumber\\
&=&\sup_{k>l>0,(\alpha_q,\alpha_{q+1},\ldots,\alpha_p)>0}\frac{\sum_{d=q}^{p}\alpha_d k^{d+1}(1+\theta)}{\sum_{d=q}^{p}\alpha_d k^{d+1}(1+\theta)+(\sum_{d=q}^{p}\alpha_d l^{d+1}-\sum_{d=q}^{p}\alpha_d lk^{d})\theta}\nonumber\\
&=&\sup_{k>l>0,(\alpha_q,\alpha_{q+1},\ldots,\alpha_p)>0}\frac{\sum_{d=q}^{p}\alpha_d k^{d+1}(1+\theta)}{\sum_{d=q}^p \alpha_d\left(k^{d+1}(1+\theta)+(l^{d+1}-lk^{d})\theta\right)}\nonumber\\
&=&\max_{d\in [p]\setminus [q-1]}\sup_{k>l>0}\frac{k^{d+1}(1+\theta)}{k^{d+1}(1+\theta)+(l^{d+1}-lk^{d})\theta}\nonumber\\
&=&\max_{d\in [p]\setminus [q-1]}\sup_{t>1}\frac{t^{d+1}(1+\theta)}{t^{d+1}(1+\theta)+(1-t^{d})\theta}\nonumber\\
&=&\max_{d\in [p]\setminus [q-1]}\sup_{t>1}\frac{t^{p+1}(1+\theta)}{t^{p+1}(1+\theta)+(t^{p-d}-t^{p})\theta}\nonumber\\
&=&\sup_{t>1}\frac{t^{p+1}(1+\theta)}{t^{p+1}(1+\theta)-t^{p}\theta+\theta}.
\end{eqnarray}
\subsection{Proof of Remark \ref{fact_gen_dis}}
By the tightness of the bounds provided in Corollary \ref{cor1}, we immediately have that 
\begin{equation}\label{ineq1_fact}
\gamma([\mathcal{F}]_H)\leq \gamma_\infty([\mathcal{F}]_H).
\end{equation}
Indeed, both $\gamma([\mathcal{F}]_H)$ and $\gamma_\infty([\mathcal{F}]_H)$ are upper bounds on the Price of Anarchy of congestion games with latency functions in $[\mathcal{F}]_H$, but $\gamma([\mathcal{F}]_H)$ is tight. Furthermore, by construction of $\gamma_\theta([\mathcal{F}]_H)$, $\eta_\theta([\mathcal{F}]_H)$, and $\gamma_\infty([\mathcal{F}]_H)$, we have that $\eta_\theta([\mathcal{F}]_H)\leq \gamma_\theta([\mathcal{F}]_H)\leq \gamma_\infty([\mathcal{F}]_H)$ for any $\theta\geq 0$, thus  
\begin{equation}\label{ineq2_fact}
\lim_{\theta\rightarrow \infty}\eta_\theta([\mathcal{F}]_H)\leq \lim_{\theta\rightarrow \infty}\gamma_\theta([\mathcal{F}]_H)\leq \gamma_\infty([\mathcal{F}]_H). 
\end{equation}

By \eqref{ineq1_fact} and \eqref{ineq2_fact}, to prove the claim it is sufficient showing that $\lim_{\theta\rightarrow \infty}\eta_\theta([\mathcal{F}]_H)\geq \gamma_\infty([\mathcal{F}]_H)$. To this aim, we show that, for any fixed $M<\gamma_\infty([\mathcal{F}]_H)$, there exists $\theta>0$ such that $\eta_\theta([\mathcal{F}]_H)>M$. Fix $M>0$. By construction of $\gamma_\infty([\mathcal{F}]_H)$, there exist $k,l,f$ such that $f\in [\mathcal{F}]_H$, $0<l<k$, and $\frac{kf(k)}{f(k)(k-l)+lf(l)}>M$. Then, there exists a sufficiently large $\theta>0$ such that $M<\frac{kf(k)+kf(k)\theta}{kf(k)+[f(k)(k-l)+lf(l)]\theta}\leq \eta_\theta([\mathcal{F}]_H)$, and this shows the claim.
\end{APPENDIX}
\theendnotes
\section*{Acknowledgements}
F. Benita  would like to acknowledge Ministry of Education, Singapore Grant SGPCTRS1804. B. Monnot   acknowledges the SUTD Presidential Graduate Fellowship. G. Piliouras gratefully acknowledges AcRF Tier 2 grant 2016-T2-1-170, grant PIE-SGP-AI-2020-01, NRF2019-NRF-ANR095 ALIAS grant and NRF 2018 Fellowship NRF-NRFF2018-07. C. Vinci would like to acknowledge the Italian MIUR PRIN 2017 Project ALGADIMAR ``Algorithms, Games, and Digital Markets''.

\bibliographystyle{oega}
\bibliography{sample-bibliography,old}
\end{document}